\documentclass[%
 reprint,
 amsmath,amssymb,
 aps,
]{revtex4-1}

\usepackage{float}
\usepackage{graphicx}
\graphicspath{ {images/} }
\usepackage{dcolumn}
\usepackage{bm}
\usepackage{url}
\usepackage{amsmath}
\usepackage{amssymb}
\usepackage{multirow}
\usepackage{array}
\usepackage{ulem}

\usepackage[usenames,dvipsnames]{color}

\usepackage{soul}

\begin{document}


\title{Intrinsic inverse band gap versus polarization relation in ferroelectric materials}

\author{Nicol\'as Forero-Correa$^{(1,2)}$, Nicolas Varas-Salinas$^{(3,4)}$, Tom\'as M. Castillo$^{(3,5)}$ and Sebastian E. Reyes-Lillo$^{(3)}$}
\email{sebastian.reyes@unab.cl}
\affiliation{
(1) Doctorado en Fisicoqu\'imica Molecular, Universidad Andres Bello, Santiago 837-0136, Chile  \\
(2) Departamento de Ciencias F\'isicas, Universidad de La Frontera, Temuco 481-1230, Chile \\
(3) Departamento de F\'isica y Astronom\'ia, Universidad Andres Bello, Santiago 837-0136, Chile\\
(4) Departamento de F\'isica, Universidad T\'ecnica Federico Santa Mar\'ia, Casilla 110-V, Valpara\'iso, Chile \\
(5) Instituto de F\'isica, Pontificia Universidad Cat\'olica de Chile, Santiago 833-1150, Chile
}\date{\today}

\begin{abstract}
Ferroelectric materials have promising applications in solar-energy conversion and electro-optic devices. The internal gradient fields produced by the macroscopic polarization may improve electronic and transport semiconducting properties. However, ferroelectrics tend to display relatively large band gaps and hence low solar-energy conversion efficiencies. In this work, we explore materials with an intrinsic inverse relation between band gap and polarization in a single ferroelectric phase. Ferroelectrics with an inverse band gap versus polarization relation are characterized by low density of states contribution at the conduction states and negligible orbital hybridization at the valence states. We use high-throughput and first principles methods to find 15 ferroelectric materials with an inverse band gap versus polarization relation in the Materials Project database. Our work provides a new pathway to design small-band gap large-polarization ferroelectrics, by simultaneously tailoring the band gap and polarization of ferroelectrics with an inverse relation through an external tuning parameter.
\end{abstract}

\pacs{Valid PACS appear here}
\maketitle

New functional properties are desired to identify and optimize materials for solar-energy conversion applications~\cite{You2018, Tyunina2015, Abel2013}. Ferroelectrics, dielectric materials with a spontaneous and switchable macroscopic polarization, have promising potential applications in photovoltaic~\cite{Butler2015,Qin2008}, photocatalytic~\cite{Li2014} and electro-optic devices~\cite{Scott2007,Sando2018, Hu2009}. Previous work has suggested that the internal gradient field produced by the ferroelectric polarization may improve carrier transport~\cite{Han2022, Tan2022}, electron-hole separation~\cite{Zheng2015}, optical properties~\cite{Wessels2007}, and photocurrent efficiencies~\cite{Young2012, Dai2023, Daranciang2012}, sparking great interest in the electronic properties of ferroelectrics and their properties under sunlight irradiation~\cite{Kreisel2012,Seidel2014,Kim2018}.

However, ferroelectrics tend to have relatively large band gaps ($\sim3-4$~eV)~\cite{Li2017c,Choi2012} and are therefore normally unsuitable as photoactive semiconductors. Hence, previous work has focused on the discovery of so-called small band gap ferroelectrics ($\sim1-2$~eV)~\cite{Bennett2012}, and the development of semiconducting ferroelectrics for electro-optic devices~\cite{Guo2013a} and solar-energy conversion applications~\cite{Han2022,Grinberg2020,Yang2012}. Despite great progress, the list of small band gap ferroelectrics remains limited~\cite{Bennett2012d, Wang2015b}. Additionally, experimental verification of small band gap ferroelectrics predicted by theory is hindered by charge leakage and sample stability~\cite{Zhang2017,Wang2014a}.

Previous work on ferroelectrics has shown that the electronic band gap and macroscopic polarization can be tuned using hydrostatic pressure~\cite{Ghosh2021,Bonomi2018}, epitaxial strain~\cite{Yang2016, Berger2011, Vonruti2018, Manzeli2015}, superlattices~\cite{Zhang2017}, quantum confinement~\cite{Reyes-Lillo2016, Birol2013}, alloying~\cite{Shimada2011, Qi2011}, chemical composition~\cite{Islam2021, Wang2022a, Choi2012a}, charge disproportion~\cite{He2017} and structural transitions~\cite{Wang2014a}. 

For the case of the prototypical perovskite SrTiO$_3$, the application of biaxial epitaxial strain on the non-polar cubic structure induces a band gap reduction due to the splitting of the t$_{2g}$ Ti-$d$ conduction edge states~\cite{Berger2011}. However, polar off-centering of the Ti ions and antiferrodistortive octahedral rotations have the opposite effect and induce a band gap increase. The band gap increase arises due to the increasing orbital hybridization between Ti-$d$ and O-$p$ states and the subsequent repulsion between the edge states forming the band gap~\cite{Berger2011, Qi2011}. These results suggest a competition between symmetry breaking and bond length, with opposite effects on the electronic band gap. While symmetry breaking of orbital degeneracies tends to narrow the band gap, structural distortions leading to smaller bond lengths and increasing orbital hybridization tend to increase the band gap. In the case of SrTiO$_3$, the latter effect dominates over the former, and results in a band gap increase with ferroelectric polarization increase, i.e. band gap and polarization display a direct relation. Similar direct relations between band gap and polarization have been observed in other ferroelectric families~\cite{Vonruti2018,Parker2011}. More generally, a phenomenological modeling suggests a direct relation between band gap and polarization for proper ferroelectrics~\cite{Fridkinbook}.

\begin{figure}
\includegraphics[width=\columnwidth]{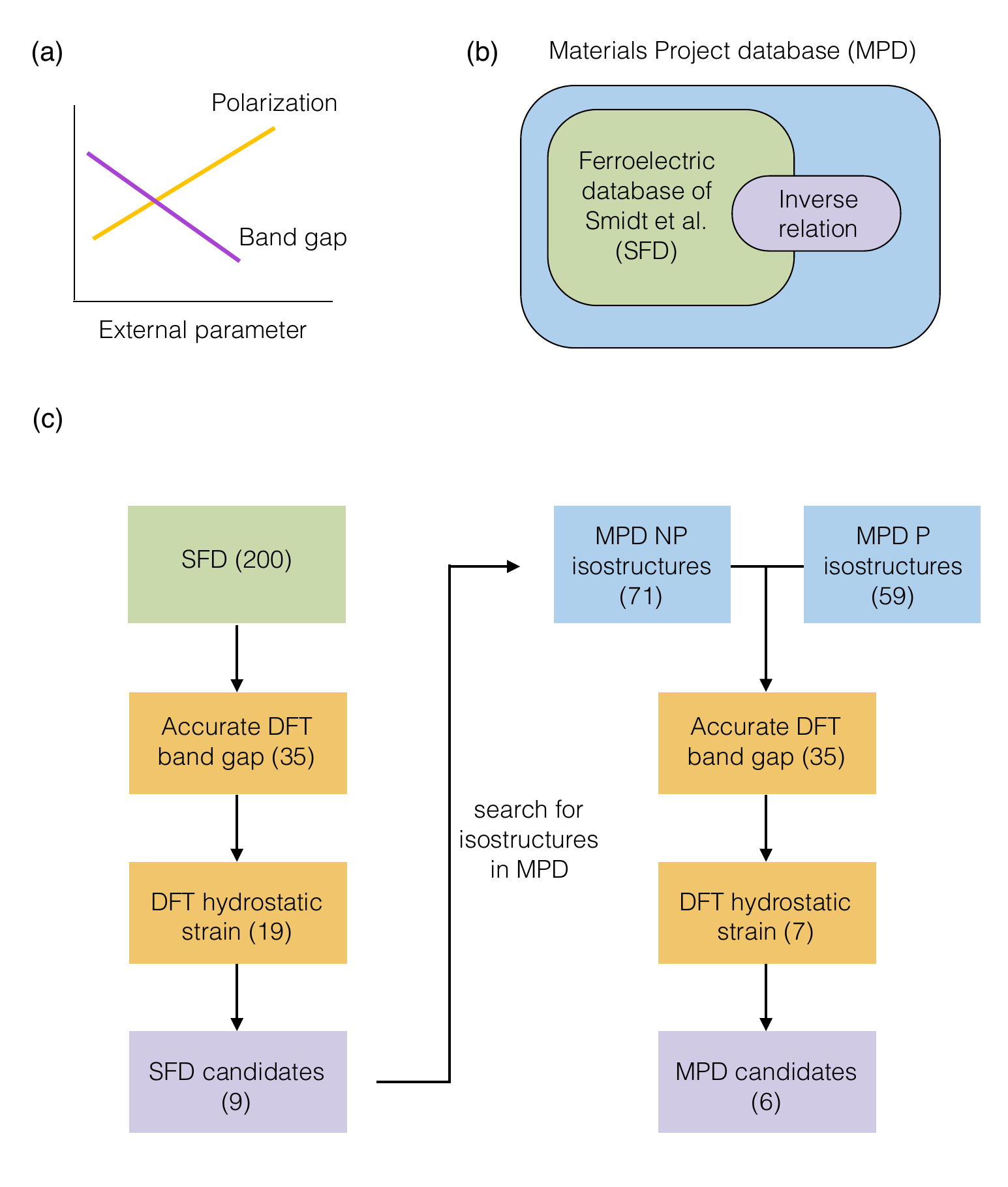}
\centering
\caption{(a) Inverse relation between polarization and band gap. (b) Schematic diagram showing the relation between the set of materials displaying an inverse band gap polarization relation, the ferroelectric database of Smidt et al. (SFD) and the Materials Project database (MPD). (c) Workflow with the relevant steps of the work and the number of materials considered in each step in parenthesis.}
\label{figure1}
\end{figure}

Nevertheless, as shown below, the microscopic interplay between the atomic polar displacement responsible for the ferroelectric polarization and the near edge state orbitals forming the band gap displays different behaviors across different families of ferroelectrics. Ferroelectric mechanisms where the band gap orbitals are not associated to the polar atomic displacement could, in principle, lead to a situation where band gap and polarization are uncorrelated or have an inverse relation under an external effect, as shown in Fig.~\ref{figure1}(a). Ferroelectrics with an inverse band gap polarization relation may lead to new pathways to search for small band gap ferroelectrics, and more generally, new strategies to simultaneously optimize band gap and polarization values for solar-energy applications through external effects such as chemical doping~\cite{Ma2020,Choi2012a}, strain~\cite{Rus2016,Sando2018}, or pressure~\cite{Ghosh2021} engineering. Notably, a concomitant band gap reduction and polarization increase has been observed due to phase transitions and film thickness~\cite{Tyunina2015, Sando2018}. Here, we explore the possibility of a simultaneous band gap decrease and polarization enhancement within a single ferroelectric phase.

To explore the possibility of such property in ferroelectrics, we must investigate a wide variety of microscopic ferroelectric mechanisms in different families of ferroelectrics. Recently, materials databases of ferroelectric materials have been constructed by integrating symmetry analysis, high throughput methods and first principles calculations~\cite{Smidt2020, Ricci2024}. The ferroelectric database of Smidt et al. (SFD)~\cite{Smidt2020} reports a dataset of 200 high quality ferroelectric materials. Ferroelectric materials are identified in the Materials Project database (MPD)~\cite{Jain2013} by searching for materials possessing a non-polar and a polar structure related through a continuous symmetry breaking, and computing the Berry phase polarization along an adiabatic insulating path connecting the non-polar and polar structures. For each ferroelectric material identified, the structural ferroelectric mechanism is provided, namely the high-symmetry non-polar reference structure, the low-symmetry polar structure, and the symmetry distortion connecting the high and low symmetry structures. The ferroelectric mechanism represents a class or family of ferroelectric materials, and here we assume that each ferroelectric compound reported in the SFD corresponds to a representative of the ferroelectric class. 

In this work, we use a combination of high-throughput and first-principles methods to search for ferroelectric materials displaying an inverse relation between polarization and electronic band gap under hydrostatic strain. We propose this intrinsic property as a new strategy to design small band gap ferroelectrics. To our knowledge, such a novel property has not been reported previously. Here we report 15 ferroelectrics with an inverse band gap versus polarization relation among different families of ferroelectrics (Fig.~\ref{figure1}(b)). We first identify ferroelectrics with an inverse relation within the SFD and later generalize our search to identify new ferroelectrics with inverse relation within the MPD.

High-throughput selection of materials is performed using the Materials Project. Materials properties are explored using \texttt{Pymatgen}~\cite{Jain2011,Ong2013}. Density functional theory (DFT) calculations are performed using the Vienna Ab initio Simulation Package (\texttt{VASP}) code~\cite{Kresse1996a,Kresse1996}. We use the generalized gradient approximation of Perdew, Burke, Ernzerhof (PBE)~\cite{Perdew1996}. In selected cases, we use the Hubbard U approximation to properly describe localized $d$ orbitals of transition metal ions~\cite{Liechtenstein1995,Dudarev1998}. Our calculations use the Materials Project settings, including an energy cut-off of 520~eV, symmetry adapted $\mathbf{k}$-point grids, projected augmented wave pseudopotentials~\cite{Joubert1999a} and Hubbard U parameters~\cite{Wang2006a, Jain2011a}. Density of states are computed using dense grids (12 $\mathbf{k}$-points per 2$\pi$ 0.25 \AA$^{-1}$ reciprocal space length). Band structures are plotted along symmetry lines with \texttt{Pyprocar}~\cite{Hinuma2017, Herath2020}. Selected materials are studied using the Heyd, Scuseria, Ernzerhof (HSE) hybrid functional~\cite{Heyd2003,Heyd2006}. Structural relaxations are performed until forces are smaller than 0.001~eV/\AA. Irreducible representation of polar modes are obtained with \texttt{Phonopy}~\cite{Togo2015}. Macroscopic polarization is computed using the modern theory of polarization, as implemented in \texttt{VASP}~\cite{King-Smith1993}.

Fig.~\ref{figure1}(c) shows the workflow of our work. We start by exploring the possibility of ferroelectrics with an inverse band gap polarization relation within the SFD. The SFD reports the structural information of the non-polar and polar structures, and is, therefore suitable for the screening strategy proposed here. Our search strategy can be generalized to other materials databases, including the ICSD~\cite{Zagorac2019}, NOMAD~\cite{Scheidgen2023}, JARVIS~\cite{Wines2023}, and Materials Cloud~\cite{Talirz2020}, to name a few. We leave the study of other databases for future work. As a first step, we identify the set of materials in the SFD having a polar band gap $E^{P}_{g}$ smaller than the reference non-polar band gap $E^{NP}_{g}$, i.e. $E^{P}_{g} < E^{NP}_{g}$. The motivation behind this criteria is as follows. The polarization is expected to monotonically increase along the adiabatic path connecting the non-polar and polar structures. Therefore, if the band gap decreases along the same adiabatic path, any external parameter that further enhances the polar distortion will simultaneously induce a band gap reduction. Therefore, the criteria $E^{P}_{g} < E^{NP}_{g}$ is expected to suggest an inverse relation between band gap and polarization. 

\begin{figure*}
\includegraphics[width=18cm]{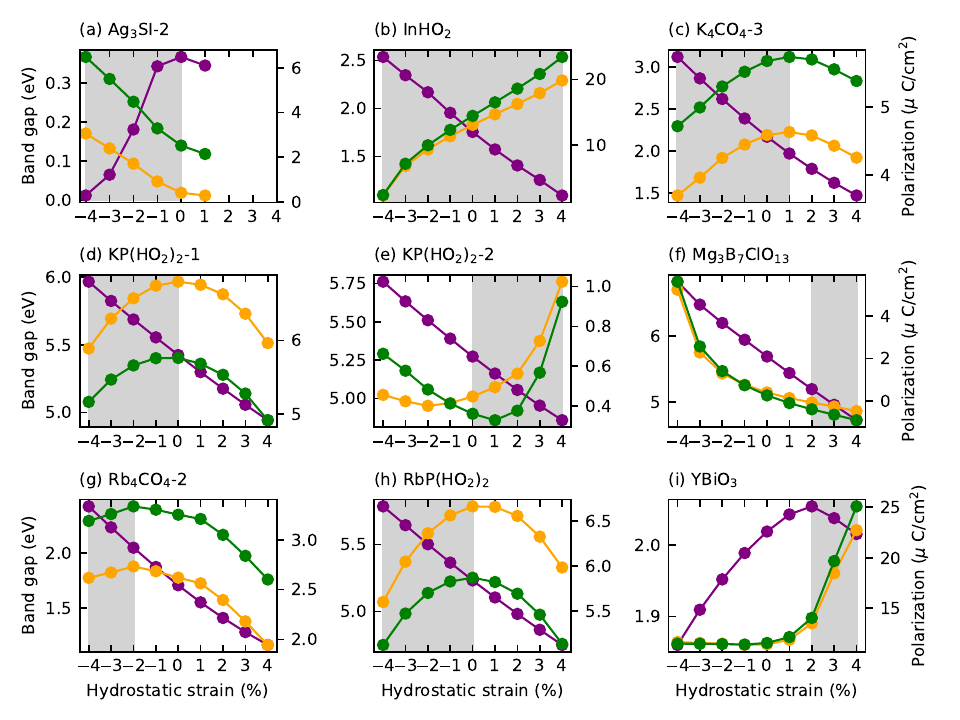}
\centering
\caption{Band gap (eV; purple) and polarization values computed with Berry phase ($\mu C/cm^2$; orange) and Born effective charges ($\mu C/cm^2$; green) as a function of hydrostatic strain. Regions with inverse band gap polarization relation are highlighted in gray.}
\label{figure2}
\end{figure*}

As a first approximation, we consider the non-polar and polar band gaps reported by the Materials Project, denoted here as $E^{NP}_{g, MPD}$ and $E^{P}_{g, MPD}$ respectively, and compute $\Delta E_{g, MPD} = E^{NP}_{g, MPD}-E^{P}_{g, MPD}$ for the materials in the SFD. From the initial 200 materials in SFD, we find that 62 ferroelectric materials (31\%) display a band gap reduction from the non-polar to the polar structure, i.e.  $\Delta E_{g, MPD}>0$. Among these, 15 cases are reported with polarization values $<0.1\mu$C/cm$^2$, too small to investigate polarization trends under an external parameter, and are therefore disregarded. In addition, 5 cases display a large volume increase $\Delta V/V>10$\% between the non-polar and polar structure. Here, we consider these volume expansions rather unphysical and we therefore disregard these cases. We further eliminate 7 materials containing only reactive non-metal ions (e.g. H$_2$O) since these typically correspond to specific structural transitions and are unlikely to represent a family of ferroelectrics. 

Table S1 reports the remaining 35 ferroelectric materials with $\Delta E_{g, MPD}>0$. Note that some cases have the equal stoichiometry but different ferroelectric mechanisms (e.g. Ag$_3$SI), because either the nonpolar, the polar or both phases are different. These cases are labeled with a number after the stoichiometry (e.g. Ag$_3$SI-1 and Ag$_3$SI-2). To verify the band gap values reported by the Materials Project, we perform full structural relaxations and compute band gaps using dense-grid density of states calculations. Table S1 compares band gap values reported in the Materials Project with our computed band gaps. Our results are generally in line with the Materials Project. In a few cases, small band gap differences lead to qualitative differences between $\Delta E_{g,MPD}$ and $\Delta E_{g}$. These cases are explained by initially small band gap differences between non-polar and polar structures, and by slightly different numerical input parameters in the calculations. We disregard 7 cases with band gap difference below $<0.01$~eV, where band gap changes are expected to be negligible. We confirm 19 cases with a band gap reduction from the non-polar to polar structure, i.e $\Delta E_{g}>0$ (highlighted in green in Table S1).

\begin{table*}[htp]
\caption{For each ferroelectric identified by our workflow, we report the space group symmetry of the non-polar and polar structures, the irreducible representation (irrep.) and frequency of the lowest frequency polar mode, macroscopic polarization $P$ computed with Berry phase and Born effective charges approximation, PBE and HSE band gaps $E_g$, and orbital character of the conduction band minimum (CBM) and valence band maximum (VBM). Experimentally references are included next to the space group symmetry when available (alphabetical order).}
\begin{tabular}{l lllr rrr cc}
\hline \hline
Prototypical & Non polar  & Polar  & Irrep. & $\omega_{FE}$  &$P_{Berry}$    & $E^{PBE}_g$  & $E^{HSE}_g$   & CBM & VBM \\
 material        & sp. group & sp. group       & polar mode  &($cm^{-1}$)    &($\mu C/cm^2$) &  (eV) &  (eV) &   & \\ \hline

Ag$_{3}$SI-2          & \textsl{Pm$\bar{3}$m}~\cite{Didisheim1986}  &  \textsl{P2$_{1}$}   & $\Gamma_4^-$(Ag,S,I) & 56.1i  &0.4     & 0.34 & 1.22  & Ag-$s$/S-$s$ & S-$p$/I-$p$ \\

InHO$_{2}$          & \textsl{Pnnm}~\cite{Christensen1964}          &  \textsl{Pmn2$_{1}$}~\cite{Lehmann1970} &  $\Gamma_2^-$(In,H,O) & 442.3i &13.1      & 1.76 & 3.57  & In-$s$ & O-$p$ \\

K$_4$CO$_4$-3 & \textsl{P$\bar{4}$3m} & \textsl{R3m}& $\Gamma_5$(K,C,O) & 72.6 & 4.6 &2.17&3.97& K-$s$& O-$p$\\

KP(HO$_{2}$)$_{2}$-1  & \textsl{I$\bar{4}$2d}~\cite{Endo1989}  &  \textsl{Fdd2}~\cite{Miyoshi2011}       & $\Gamma_4$(P,H,O) & 516.9i  &0.6      & 5.42 & 7.33 & K-$s$/P-$s$/O-$s$ & O-$p$ \\

KP(HO$_{2}$)$_{2}$-2  & \textsl{P2$_1$/c}  &  \textsl{Cc} &  $\Gamma_1^-$(K,P,H,O) & 494.5i &0.4 & 5.28 & 6.98 &K-$s$/O-$s$& O-$p$\\

Mg$_{3}$B$_{7}$ClO$_{13}$ & \textsl{F$\bar{4}$3c}~\cite{Sueno1973} & \textsl{Pca2$_{1}$}~\cite{Wang2018} &  $\Gamma_5$(Mg,B,Cl,O)& 115.1i &0.4   &   5.69 & 7.73 & Mg-$s$/Cl-$s$ & O-$p$ \\

Rb$_{4}$CO$_{4}$-2    & \textsl{P$\bar{4}$3m}~\cite{Mattauch2004}  &  \textsl{R3m}        & $\Gamma_5$(Rb,C,O) &  49.1 & 2.6  & 1.83 & 3.37  & Rb-$s$ & O-$p$ \\

RbP(HO$_{2}$)$_{2}$  & \textsl{I$\bar{4}$2d}  &  \textsl{Fdd2}&$\Gamma_4$(P,H,O) & 532.4i &6.7& 5.23 & 7.09 & Rb-$s$/P-$s$/O-$s$& O-$p$\\

YBiO$_{3}$          & \textsl{P6$_{3}$/mmc}   &  \textsl{P6$_{3}$cm} &  $\Gamma_2^-$(Y,Bi,O) & 100.3i & 11.4 & 2.02 & 2.83   & Bi-$p$/O-$p$ & O-$p$ \\

\hline \hline  
\end{tabular}
\label{lista1}
\end{table*}

Next, for the 19 cases displaying a band gap reduction, we construct an adiabatic structural path connecting the non-polar and polar structures to compute the ferroelectric polarization. Table S2 reports our results for total energy differences, macroscopic polarizations, and volume changes along the adiabatic path connecting the non-polar and polar structures. Our computed polarizations using the Berry phase approach are in overall good qualitative agreement with the values reported in the SFD. Quantitative differences are explained by the stringent force threshold used here for relaxation. In some cases, non-polar reference structures correspond to shallow local minima and require small distortion steps to remain within the local minimum. Energies above hull for the polar structures are extracted from the Materials Project.

In 4 cases, namely Mg(NO$_3$)$_2$, Mo(HO$_2$)$_2$-1, Mo(HO$_2$)$_2$-2 and Zn(NO$_3$)$_2$, the polar structure corresponds to a metastable phase. In all these cases, the MPD reports the non-polar structure as the ground state structure, in agreement with our results. Notably, Mg(NO$_3$)$_2$ and Zn(NO$_3$)$_2$ belong to the same family of ferroelectrics, i.e. display the same ferroelectric mechanism. The small energy difference between nonpolar and polar structures suggests the possibility of an antiferroelectric switching.

Figs. S1 and S2 show the band gap and polarization along the linearly interpolated structural path connecting the non-polar and polar structures for the materials reported in Table S2. Plots are separated into two sets for later convenience. In most cases, band gaps display a monotonic decrease as the polarization monotonically increases along the structural path connecting non-polar and polar structures. The latter suggests that an inverse relation between band gap and polarization may arise if the polar distortion is further enhanced through some external parameter. In few cases, band gaps display non-smooth or non-monotonic behavior along the adiabatic path due to changes in band gap position in the Brillouin zone. For instance, for Ag$_3$SI-2, the band gap changes from a direct gap at $\Gamma$ to an indirect gap between the segment B-$\Gamma$ and $\Gamma$. In other cases, the band gap changes positions due to the large rearrangements of the ions along the adiabatic path (e.g. Mg(NO$_3$)$_2$, Mo(HO$_2$)$_2$-1, YScO$_3$ in Fig. S2). Polarization values obtained with the Berry phase approach are corroborated using the linear approximation $P \simeq e/V \sum_i Z_i \delta u_i$, where $Z_i$ and $\delta u_i$ are the Born effective charges and atomic displacement between non-polar and polar structures for atom $i$, respectively, $V$ is the polar structure volume, $e$ is the electron charge, and the sum runs over all the ions in the structure. In both cases, we use the non-polar structure as a reference. Therefore, as expected, polarization values obtained with Born effective charges display a linear behavior and agree with the Berry phase approach for small polar distortions. 

In the following, we investigate the relation between band gap and polarization under the effect of hydrostatic strain for the 19 materials in Table S2. Here, hydrostatic strain is chosen as a simple way to increase the volume of the structure without changing the symmetry of the ferroelectric phase. Indeed, we find that 9 ferroelectrics display an inverse relation between band gap and polarization (highlighted in green in Table S2), whereas 10 display a direct relation. Figs.~\ref{figure2} and S3 show the band gap and polarization as a function of applied hydrostatic strain for the ferroelectrics displaying an inverse and direct relation, respectively. Regions of inverse band gap and polarization relation are highlighted in gray. Ag$_3$SI-2 becomes unstable above 1\% strain and undergoes a structural distortion; we therefore disregard that portion of the graph. The same occurs for Mo(HO$_2$)$_2$-1 and Zn(NO$_3$)$_2$ at compressive and tensile strain, respectively (see Fig. S3).

\begin{figure*}
\includegraphics[width=18cm]{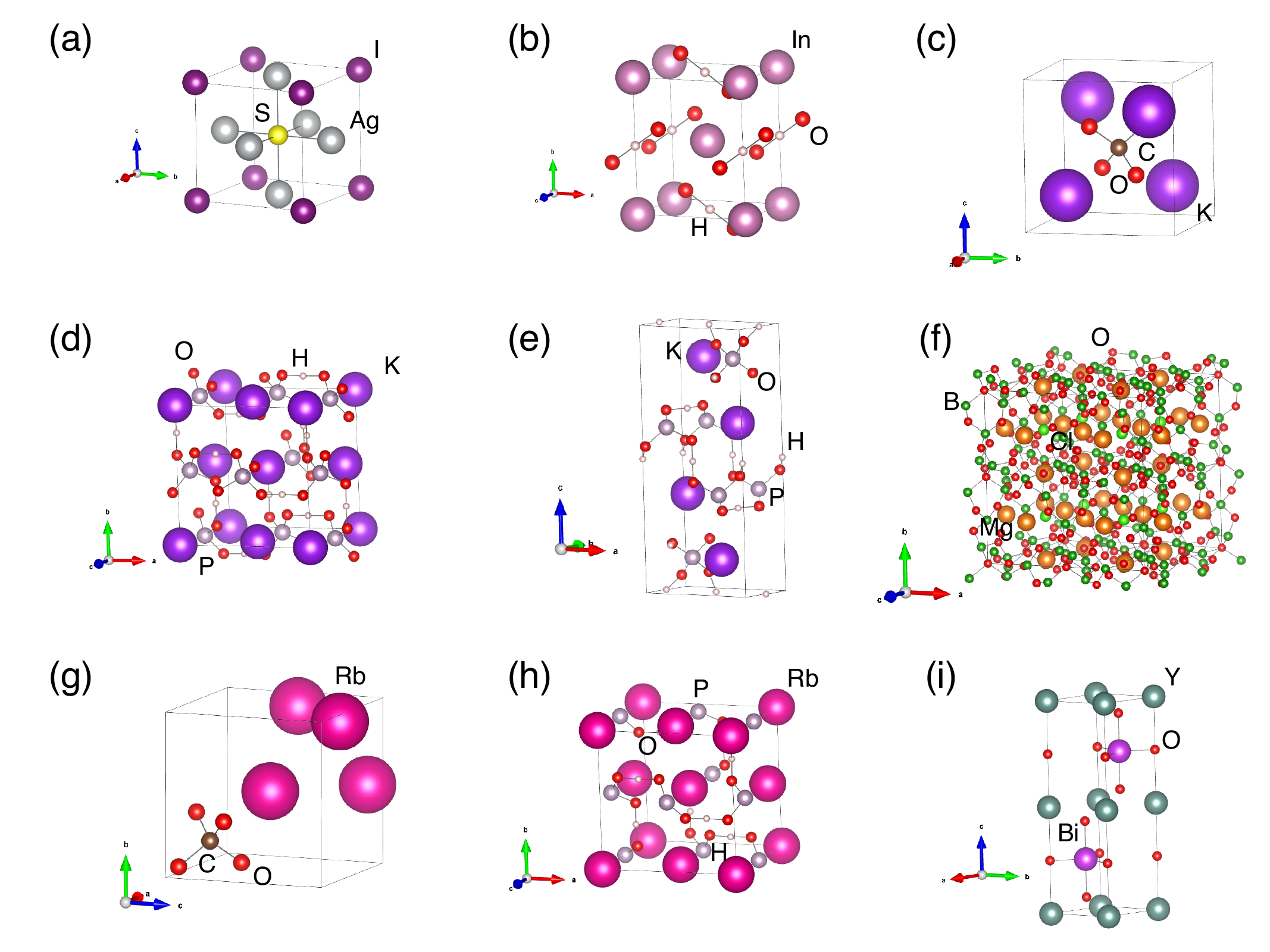}
\centering
\caption{Atomic structure for the ferroelectric materials reported in Table~\ref{lista1}: (a) Ag$_{3}$SI-2, (b) InHO$_{2}$, (c) K$_4$CO$_4$-3, (d) KP(HO$_{2}$)$_{2}$-1, (e) KP(HO$_{2}$)$_{2}$-2, (f) Mg$_{3}$B$_{7}$ClO$_{13}$, (g) Rb$_4$CO$_4$-2, (h) RbP(HO$_{2}$)$_{2}$ and (i) YBiO$_{3}$. Structures are plotted with VESTA~\cite{Momma2008}.}
\label{figure3}
\end{figure*}

For Ag$_3$SI-2 and YBiO$_3$ the band gap decreases at large compressive strain, with decreasing volume, and the structures become metallic. For the rest of the materials in Fig.~\ref{figure2}, the band gap decreases monotonically with increasing strain. In these cases, the band gap is modulated by the volume, not by the magnitude of the polarization, and decreases due to the increasing bond lengths in the structure. In parallel, the polarization has a monotonic behavior for Ag$_3$SI-2, InHO$_2$ and YBiO$_3$. For the case of Mg$_3$B$_7$ClO$_{13}$, the polarization changes sign around 2\%, and the polarization magnitude increases above 2\% strain. YBiO$_3$ displays a nearly constant polarization for compressive strain and a sharp increase above 1\%. For K$_4$CO$_4$-3, KP(HO$_{2}$)$_{2}$-1, KP(HO$_{2}$)$_{2}$-2, Rb$_4$CO$_4$-2 and RbP(HO$_{2}$)$_{2}$ the polarization displays a non-monotonic behavior. The polarization reaches a maximum (minimum) value near 0\% strain for K$_4$CO$_4$-3, KP(HO$_{2}$)$_{2}$-1, Rb$_4$CO$_4$-2 and RbP(HO$_{2}$)$_{2}$ (for KP(HO$_{2}$)$_{2}$-2), and decreases (increases) at both compressive and tensile strains. Table S5 reports the band gap and polarization percentage variation as a function of hydrostatic strain around the linear part of the plots in Fig.~\ref{figure2}. Our results show that Ag$_3$SI-2 displays the largest simultaneous inverse band gap and polarization variation among the candidates.

Fig.~\ref{figure3} shows the atomic structure for the 9 ferroelectrics displaying an inverse band gap polarization relation. Table~\ref{lista1} reports their space group symmetries, irreducible representations and frequency of the lowest ferroelectric polar mode, polarization, as well as band gaps and band edge orbital character of the conduction band minimum (CBM) and valence band maximum (VBM). KP(HO$_2$)$_2$-1, Mg$_3$B$_7$ClO$_{13}$, KP(HO$_{2}$)$_{2}$-1 and RbP(HO$_{2}$)$_{2}$ are well-known ferroelectrics~\cite{linesbook}. Ag$_3$SI-2, InHO$_2$ and Rb$_4$CO$_4$-2 have experimentally accessible structures. The rest of the materials have been theoretically predicted with first principles~\cite{Cancarevic2007}. 

Polar modes are obtained by performing a $\Gamma$-centered phonon mode calculation for the non-polar structure. From the 9 candidate materials, 7 display a polar mode with imaginary frequencies and therefore correspond to proper ferroelectrics. The 2 other cases, namely K$_4$CO$_4$-3 \textsl{P$\bar{4}$3m} and Rb$_4$CO$_4$-2 \textsl{P$\bar{4}$3m} display real frequencies, signaling an improper ferroelectric mechanism. These results show that an inverse relation between band gap and polarization is not exclusive for proper or improper ferroelectric mechanisms, but rather an intrinsic property of the polar phase.

Semi-local functionals such as PBE tend to underestimate experimental band gaps due to well-known limitations of DFT~\cite{Mori-Sanchez2008}. Therefore, to confirm the band gap narrowing, we perform full structural relaxations using HSE and compute HSE band gaps with dense-grid density of states calculations. Table~\ref{lista1} reports band gaps computed with PBE and HSE. Table S3 reports HSE lattice parameters and HSE band gap differences for the materials in Table~\ref{lista1}. HSE displays a large ($\sim30$\%) band gap increase and relatively small lattice parameter change with respect to PBE. Notably, HSE confirms the band gap narrowing between the non-polar and polar structures (see Table S3). With the only exception of YBiO$_3$, spin-orbit interaction has a negligible effect on the magnitude of the polar structures band gaps ($<0.01$ eV). For YBiO$_3$, spin-orbit interaction decreases the magnitude of the band gap but maintains the trend (see Fig. S4). 

\begin{figure*}
\includegraphics[width=18cm]{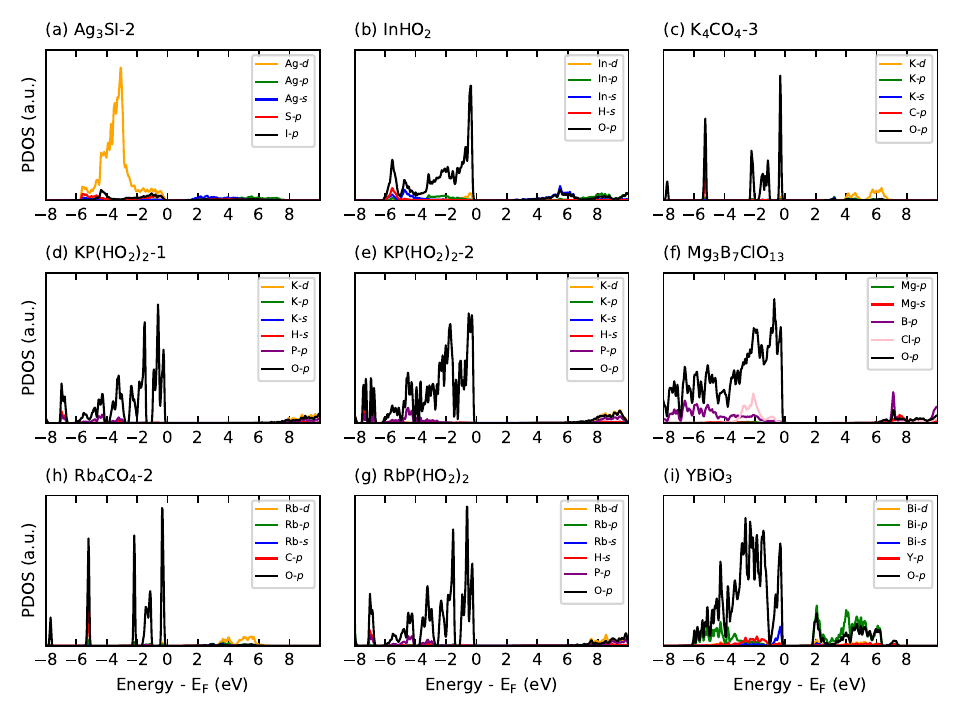}
\centering
\caption{Density of states for the 9 polar structure displaying an inverse band gap polarization relation under the effect of hydrostatic strain. The Fermi level (E$_F$) is set to zero.}
\label{figure4}
\end{figure*}

\begin{figure*}
\includegraphics[width=18cm]{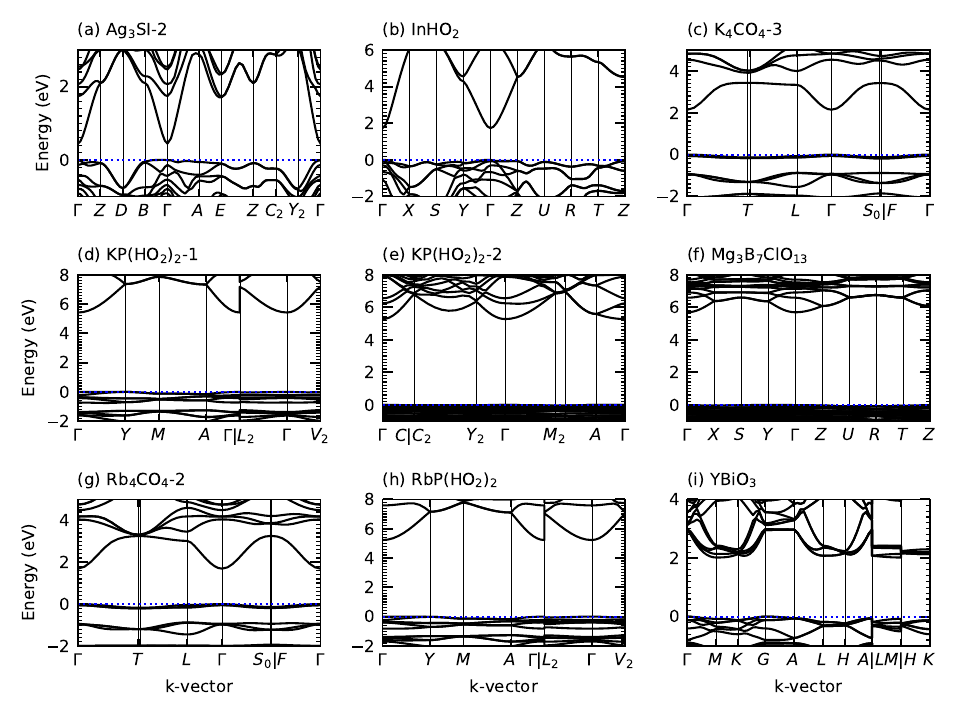}
\centering
\caption{Band structures for the 9 polar structure displaying an inverse band gap polarization relation under the effect of hydrostatic strain reported in Table~\ref{lista1}. The Fermi level (E$_F$) is set to zero.}
\label{figure5}
\end{figure*}

To explore the origin of the inverse relation between band gap and polarization, we investigate the electronic properties of the materials. Figs.~\ref{figure4} shows DFT-PBE partial density of states for the materials reported in Table~\ref{lista1}. The main common characteristic among the different densities of states is the small density contribution at the conduction states, relative to the respective valence states (similarly for some materials in Fig. S3). Accordingly, as shown in Fig.~\ref{figure5}, the band structures display dispersive conduction bands. The lack of conduction states density also explains the small effect of spin-orbit interaction in the band gap~\cite{Krach2023}. In addition, the dominating orbital contribution at the valence states displays negligible hybridization with other types of orbitals. 

For Ag$_3$SI, the main polar distortion involves displacements of S against Ag. However, Ag-$d$ and S-$p$ display negligible hybridization in the density of states. Similarly, for the sets (i) InHO$_2$, KP(HO$_2$)$_2$-1, KP(HO$_2$)$_2$-2, RbP(HO$_2$)$_2$, (ii) K$_4$CO$_4$, Rb$_4$CO$_4$-2 and (iii) Mg$_3$B$_7$ClO$_{13}$, the polar mode involves polar displacements between (i) H, O, (ii) C, O, and (iii) B, O, respectively, which hybridizes deep into the valence states, and therefore has negligible effect on band gap. Finally, the improper ferroelectric mechanism of YBiO$_3$ involves displacements of Y against O. However, Y-$p$ and O-$p$ display negligible hybridization in the density of states.

The lack of density contribution at the conduction states and hybridization in the valence states leads to a small interaction between the near edge gap states and the polar distortion. More specifically, the atomic orbitals participating in the polar distortion have negligible contribution at the band gap edges, and therefore, polarization and band gap are effectively decoupled. The ferroelectrics in Fig.~\ref{figure2} can be described as geometric, in the sense that the polarization is not driven by changes in chemical bonding or orbital hybridization, but rather due to electrostatics and size effects. The latter explains the non-monotonic behavior of the polarization for K$_4$CO$_4$-3, KP(HO$_{2}$)$_{2}$-1, KP(HO$_{2}$)$_{2}$-2, Rb$_4$CO$_4$-2 and RbP(HO$_{2}$)$_{2}$. At large compressive strain, i.e. small volumes, atoms have little room to displace and the polarization decreases. At large tensile strain, the polar displacement reaches a maximum and the polarization decreases due to the volume increase. For Ag$_3$SI, InHO$_2$ and YBiO$_3$, the behavior of the polarization under hydrostatic strain is specific to their ferroelectric mechanism.

\begin{table}[htp]
\caption{List of ferroelectrics with either the non polar or polar phase isostructural to one of the ferroelectrics reported in Table~\ref{lista1}. For each ferroelectric we report the ferroelectric family, Berry phase polarization $P$, band gap $E_g$, as well as total energy $\Delta E$ and volume ($\Delta V/V$) difference between non-polar and polar structures (alphabetical order).}
\begin{tabular}{lcc cc }
\hline \hline
Ferroelectric   &  $P$ &  $E^{PBE}_g$   & $\Delta E$  & $\Delta V/V$ \\
material     & ($\mu C/cm^2$)&   (eV) &  (meV) & (\%)  \\ \hline
AlHO$_{2}$ (\textsl{Pmn2$_{1}$})~\cite{Komatsu2006}      & 21.0 & 5.69  & -36.9 &   2.2    \\
CsAs(HO$_2$)$_2$  (\textsl{Fdd2}) & 7.1 & 3.91 &   -65.0& 2.6  \\

GaHO$_{2}$ (\textsl{Pmn2$_{1}$})      & 20.0 & 2.98  & -46.0 &  2.4    \\
Mg$_3$B$_7$BrO$_{13}$ (\textsl{F$\bar{4}$3c})  & 1.1 & 5.69  & -95.8 &  1.1 \\

YHO$_{2}$ (\textsl{Pmn2$_{1}$})      & 17.2 & 4.84  & -127.4 &  3.8  \\

Zn$_3$B$_7$BrO$_{13}$ (\textsl{Pca2$_{1}$})  & 0.4 & 4.45  & -300.1 &  1.2  \\

\hline \hline  
\end{tabular}
\label{lista2}
\end{table}

Interestingly, KP(HO$_{2}$)$_{2}$-1 and RbP(HO$_{2}$)$_{2}$ as well as K$_4$CO$_4$-3 and Rb$_4$CO$_4$-1 belong to the same family of ferroelectrics, and YBiO$_3$ is isostructural to the well-known geometric ferroelectric YMnO$_3$~\cite{Aken2004}. Table S4 reports the structure types of the ferroelectric materials in Table I, as reported in the ICSD. The latter suggests the existence of more ferroelectric materials with the desired property within the ferroelectric families. To explore this possibility, we search for other ferroelectric materials with an inverse band gap polarization relation in the Materials Project (see Fig.~\ref{figure1}(b)). To this end, we generalize our high-throughput search by interpreting each prototypical ferroelectric material in the SFD as a representative of a family of ferroelectrics. The initial set of ferroelectric materials in the SFD is enlarged by considering all materials in the MPD possessing a phase isostructural to either the polar or the non-polar structure of one of the prototypical materials reported in Table~\ref{lista1}. The resulting group of materials includes new ferroelectric candidates, not included in the SFD. However, strictly speaking, these cases should not be considered as new entries in the SFD, since the methodology proposed in the SFD requires that both nonpolar and polar phases be present in MPD.

\begin{figure}
\includegraphics[width=8cm]{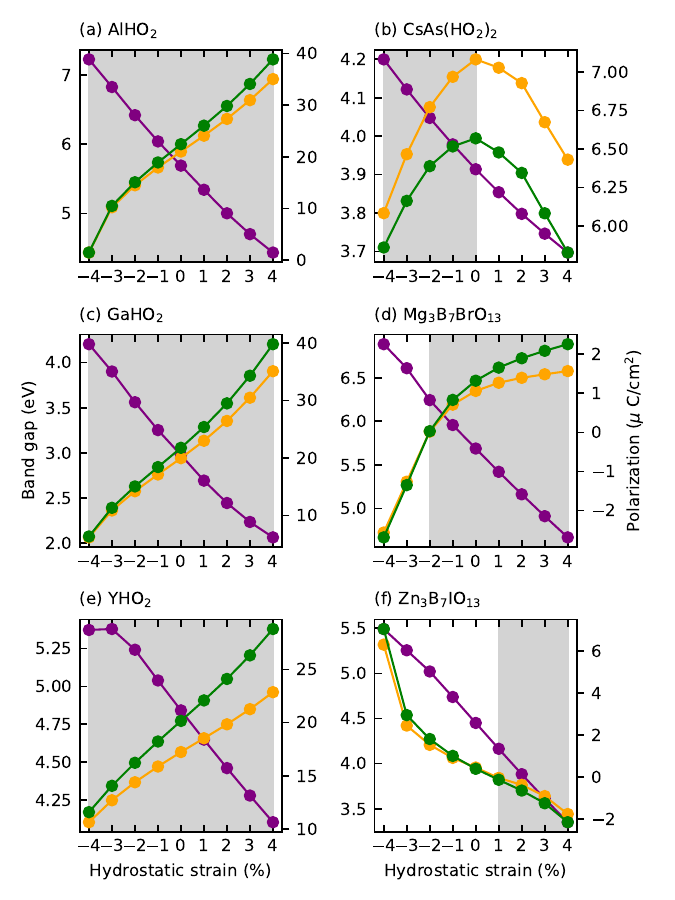}
\centering
\caption{Isostructural ferroelectrics (a) AlHO$_2$, (b) CsAs(HO$_2$)$_2$, (c) GaHO$_2$, (d) Mg$_3$B$_7$BrO$_{13}$, (e) YHO$_2$ and (f) Zn$_3$B$_7$BrO$_{13}$ displaying an indirect band gap polarization relation under the effect of isotropic strain. For each ferroelectric, we report band gap (eV; purple), Berry phase polarization ($\mu C/cm^2$; orange) and Born effective charges polarization ($\mu C/cm^2$; green)}
\label{figure6}
\end{figure}

Isostructural materials are identified within the Materials Project by imposing structural conditions, specifically, equal space group, total number of atoms, number of species, Wyckoff positions, and coordination number around each atom. These conditions are screened using the pymatgen functions~\cite{Ong2013}. Our search identifies 59 polar and 71 non-polar materials in the Materials Project isostructural to one of the prototypical materials in Table~\ref{lista2}. However, only 42 polar and 41 non-polar are insulating. From these, 20 polar and 6 non-polar cases contain lanthanides ions, whereas 12 polar and 10 non-polar cases contain magnetic ions with open $d$ orbitals. We disregard these cases since they would require magnetic calculations with multiple spin configurations and computationally intensive calculations that are out of the scope of the work. We perform full structural relaxations and compute band gaps using dense-grid density of states calculations for the remaining 10 polar and 25 non-polar materials. For the case of polar structures, the associated reference non-polar structure is constructed isostructural to the non-polar structure in the SFD. In this case, we confirm the non-polar structure is insulating, therefore allowing a direct calculation of the Berry phase polarization. Similarly, for the case of the non-polar structures, the polar structure is constructed isostructural to the polar structure in the SFD. In this case, the stability of the polar structure is determined from full structural relaxations. 

We confirm 7 cases with a band gap narrowing from non-polar to polar, namely XHO$_2$ X = Al, Ga, Y, Cd(NO$_3$)$_2$, CsAs(HO$_2$)$_2$, Mg$_3$B$_7$BrO$_{13}$ and Zn$_3$B$_7$BrO$_{13}$. From these, we find that the polar structures of XHO$_2$ X = Al, Ga, Y, CsAs(HO$_2$)$_2$, Mg$_3$B$_7$BrO$_{13}$ and Zn$_3$B$_7$BrO$_{13}$ display an inverse band gap polarization relation (see Fig.~\ref{figure6}). Table~\ref{lista2} reports their Berry phase polarization, PBE band gaps, as well as energy and volume differences between non-polar and polar phases. CsAs(HO$_2$)$_2$~\cite{linesbook} and Zn$_3$B$_7$BrO$_{13}$~\cite{Campa-Molina2006} are well-known ferroelectrics, and the polar structures of AlHO$_2$ and GaHO$_2$ are experimentally accessible. Notably, CsAs(HO$_2$)$_2$ \textsl{Fdd2} displays larger polarization and smaller PBE band gap than its isostructural counterpart KP(HO$_2$)$_2$ \textsl{Fdd2}.

We make several observations from our results summarized in Table~\ref{lista3}. We first highlight that 9 from 200 (4.5\%) prototypical ferroelectrics display an inverse relation between polarization and band gap. Families of ferroelectrics may have more than one example of ferroelectrics displaying the inverse relation property. We also note that isostructural materials within a ferroelectric family may either have or not have the inverse polarization band gap relation. These results suggest that our search is far from exhaustive.

\begin{table}
\caption{Number of materials in the ferroelectric database of Smidt et al. and the Materials Project database, materials with band gap narrowing between non-polar and polar structures, isostructural candidates, and the final candidates that exhibit an inverse band gap versus polarization relation.} 
\begin{tabular}{llr}
\hline \hline
Materials& Search      &   Number of \\ 
database & criteria    &  ferroelectrics   \\ \hline
Ferroelectrics & Total database &  200\\
database &Ferroelectrics with $E^{P}_{g} < E^{NP}_{g}$  &  19 \\
of Smidt et al.&Inverse $E_{g}$ vs. $P$ relation    &  9\\
\hline
Materials &Polar isostructures        &  59\\
Project &Non polar isostructures     &  71\\
database&Isostructures with $E^{P}_{g} < E^{NP}_{g}$  &  7 \\
&Inverse $E_{g}$ vs. $P$ relation        &  6\\
\hline \hline  
\end{tabular}
\label{lista3}
\end{table}

Interestingly, the ferroelectrics in Table~\ref{lista2} were identified as inverse band gap versus polarization ferroelectrics simply due to the fact that their prototypical ferroelectric class representative displayed the property themselves. The latter suggests that other inverse relation ferroelectrics may exist within families of ferroelectrics whose representative in the SFD does not display the property. The latter calls for a systematic search for materials with the property.

In usual ferroelectrics where the polarization arises due to chemical bonding or orbital hybridization, the behavior of the band gap is determined by polarization orientation, symmetry breakings, volume, and type of near edge orbital contribution. Here, ferroelectrics are characterized by low conduction density of states and valence orbital hybridization; therefore, band gaps and polarizations are greatly affected by the volume. We emphasize that the band gap reduction from non-polar and polar structures $\Delta E_g>0$ is not a sufficient condition to identify inverse relation ferroelectrics. In addition, not all ferroelectrics within a certain class display the inverse relation property; therefore the property is independent of the ferroelectric mechanism.  We conclude that in addition to low density of states contribution at the conduction states and negligible orbital hybridization at the valence states, inverse relation ferroelectrics require the right rates of band gap and polarization change under the external parameters, which in turn is determined by details of orbital interaction and distortions in the polar structure. 

Ferroelectrics with a small band gap and large polarization have potential applications as photovoltaics. Our results suggest that YBiO$_3$ is a promising candidate (see Table~\ref{lista1}). Table S6 shows the effective masses computed for the band structures in Fig. 5. The ferroelectric materials in Table I tend to display large (small) effective masses at the VBM (CBM), suggesting relatively poor (good) hole (electron) transport. The study of optical and transport properties of these materials is left for future work.

In summary, we have found 15 examples of ferroelectric materials displaying a novel property, an inverse relation between band gap and polarization. We find 9 examples of such type of ferroelectrics within the ferroelectric database of Smidt et al. and propose 6 candidates present in the Materials Project. Our results suggest new pathways to engineer small-band gap large-polarization ferroelectrics.

\section{\label{sec:level1} Acknowledgments}

This work was supported by ANID FONDECYT Regular grant number 1220986. N.F.-C. acknowledges partial financial support from the project InTec, Code “FRO2395”, from the Ministry of Education of Chile. Powered@NLHPC: This research was supported by the supercomputing infrastructure of the NLHPC (CCSS210001).


\begin{thebibliography}{84}%
\makeatletter
\providecommand \@ifxundefined [1]{%
 \@ifx{#1\undefined}
}%
\providecommand \@ifnum [1]{%
 \ifnum #1\expandafter \@firstoftwo
 \else \expandafter \@secondoftwo
 \fi
}%
\providecommand \@ifx [1]{%
 \ifx #1\expandafter \@firstoftwo
 \else \expandafter \@secondoftwo
 \fi
}%
\providecommand \natexlab [1]{#1}%
\providecommand \enquote  [1]{``#1''}%
\providecommand \bibnamefont  [1]{#1}%
\providecommand \bibfnamefont [1]{#1}%
\providecommand \citenamefont [1]{#1}%
\providecommand \href@noop [0]{\@secondoftwo}%
\providecommand \href [0]{\begingroup \@sanitize@url \@href}%
\providecommand \@href[1]{\@@startlink{#1}\@@href}%
\providecommand \@@href[1]{\endgroup#1\@@endlink}%
\providecommand \@sanitize@url [0]{\catcode `\\12\catcode `\$12\catcode
  `\&12\catcode `\#12\catcode `\^12\catcode `\_12\catcode `\%12\relax}%
\providecommand \@@startlink[1]{}%
\providecommand \@@endlink[0]{}%
\providecommand \url  [0]{\begingroup\@sanitize@url \@url }%
\providecommand \@url [1]{\endgroup\@href {#1}{\urlprefix }}%
\providecommand \urlprefix  [0]{URL }%
\providecommand \Eprint [0]{\href }%
\providecommand \doibase [0]{http://dx.doi.org/}%
\providecommand \selectlanguage [0]{\@gobble}%
\providecommand \bibinfo  [0]{\@secondoftwo}%
\providecommand \bibfield  [0]{\@secondoftwo}%
\providecommand \translation [1]{[#1]}%
\providecommand \BibitemOpen [0]{}%
\providecommand \bibitemStop [0]{}%
\providecommand \bibitemNoStop [0]{.\EOS\space}%
\providecommand \EOS [0]{\spacefactor3000\relax}%
\providecommand \BibitemShut  [1]{\csname bibitem#1\endcsname}%
\let\auto@bib@innerbib\@empty
\bibitem [{\citenamefont {You}\ \emph {et~al.}(2018)\citenamefont {You},
  \citenamefont {Zheng}, \citenamefont {Fang}, \citenamefont {Zhou},
  \citenamefont {Tan}, \citenamefont {Zhang}, \citenamefont {Ma}, \citenamefont
  {Schmidt}, \citenamefont {Rusydi}, \citenamefont {Wang}, \citenamefont
  {Chang}, \citenamefont {Rappe},\ and\ \citenamefont {Wang}}]{You2018}%
  \BibitemOpen
  \bibfield  {author} {\bibinfo {author} {\bibfnamefont {L.}~\bibnamefont
  {You}}, \bibinfo {author} {\bibfnamefont {F.}~\bibnamefont {Zheng}}, \bibinfo
  {author} {\bibfnamefont {L.}~\bibnamefont {Fang}}, \bibinfo {author}
  {\bibfnamefont {Y.}~\bibnamefont {Zhou}}, \bibinfo {author} {\bibfnamefont
  {L.~Z.}\ \bibnamefont {Tan}}, \bibinfo {author} {\bibfnamefont
  {Z.}~\bibnamefont {Zhang}}, \bibinfo {author} {\bibfnamefont
  {G.}~\bibnamefont {Ma}}, \bibinfo {author} {\bibfnamefont {D.}~\bibnamefont
  {Schmidt}}, \bibinfo {author} {\bibfnamefont {A.}~\bibnamefont {Rusydi}},
  \bibinfo {author} {\bibfnamefont {L.}~\bibnamefont {Wang}}, \bibinfo {author}
  {\bibfnamefont {L.}~\bibnamefont {Chang}}, \bibinfo {author} {\bibfnamefont
  {A.~M.}\ \bibnamefont {Rappe}}, \ and\ \bibinfo {author} {\bibfnamefont
  {J.}~\bibnamefont {Wang}},\ }\href {\doibase 10.1126/sciadv.aat3438}
  {\bibfield  {journal} {\bibinfo  {journal} {Science Advances}\ }\textbf
  {\bibinfo {volume} {4}},\ \bibinfo {pages} {1} (\bibinfo {year}
  {2018})}\BibitemShut {NoStop}%
\bibitem [{\citenamefont {Tyunina}\ \emph {et~al.}(2015)\citenamefont
  {Tyunina}, \citenamefont {Yao}, \citenamefont {Chvostova}, \citenamefont
  {Dejneka}, \citenamefont {Kocourek}, \citenamefont {Jelinek}, \citenamefont
  {Trepakov},\ and\ \citenamefont {{Van Dijken}}}]{Tyunina2015}%
  \BibitemOpen
  \bibfield  {author} {\bibinfo {author} {\bibfnamefont {M.}~\bibnamefont
  {Tyunina}}, \bibinfo {author} {\bibfnamefont {L.}~\bibnamefont {Yao}},
  \bibinfo {author} {\bibfnamefont {D.}~\bibnamefont {Chvostova}}, \bibinfo
  {author} {\bibfnamefont {A.}~\bibnamefont {Dejneka}}, \bibinfo {author}
  {\bibfnamefont {T.}~\bibnamefont {Kocourek}}, \bibinfo {author}
  {\bibfnamefont {M.}~\bibnamefont {Jelinek}}, \bibinfo {author} {\bibfnamefont
  {V.}~\bibnamefont {Trepakov}}, \ and\ \bibinfo {author} {\bibfnamefont
  {S.}~\bibnamefont {{Van Dijken}}},\ }\href {\doibase
  10.1088/1468-6996/16/2/026002} {\bibfield  {journal} {\bibinfo  {journal}
  {Science and Technology of Advanced Materials}\ }\textbf {\bibinfo {volume}
  {16}},\ \bibinfo {pages} {26002} (\bibinfo {year} {2015})}\BibitemShut
  {NoStop}%
\bibitem [{\citenamefont {Abel}\ \emph {et~al.}(2013)\citenamefont {Abel},
  \citenamefont {St{\"{o}}ferle}, \citenamefont {Marchiori}, \citenamefont
  {Rossel}, \citenamefont {Rossell}, \citenamefont {Erni}, \citenamefont
  {Caimi}, \citenamefont {Sousa}, \citenamefont {Chelnokov}, \citenamefont
  {Offrein},\ and\ \citenamefont {Fompeyrine}}]{Abel2013}%
  \BibitemOpen
  \bibfield  {author} {\bibinfo {author} {\bibfnamefont {S.}~\bibnamefont
  {Abel}}, \bibinfo {author} {\bibfnamefont {T.}~\bibnamefont
  {St{\"{o}}ferle}}, \bibinfo {author} {\bibfnamefont {C.}~\bibnamefont
  {Marchiori}}, \bibinfo {author} {\bibfnamefont {C.}~\bibnamefont {Rossel}},
  \bibinfo {author} {\bibfnamefont {M.~D.}\ \bibnamefont {Rossell}}, \bibinfo
  {author} {\bibfnamefont {R.}~\bibnamefont {Erni}}, \bibinfo {author}
  {\bibfnamefont {D.}~\bibnamefont {Caimi}}, \bibinfo {author} {\bibfnamefont
  {M.}~\bibnamefont {Sousa}}, \bibinfo {author} {\bibfnamefont
  {A.}~\bibnamefont {Chelnokov}}, \bibinfo {author} {\bibfnamefont {B.~J.}\
  \bibnamefont {Offrein}}, \ and\ \bibinfo {author} {\bibfnamefont
  {J.}~\bibnamefont {Fompeyrine}},\ }\href@noop {} {\bibfield  {journal}
  {\bibinfo  {journal} {Nature Communications}\ }\textbf {\bibinfo {volume}
  {4}},\ \bibinfo {pages} {1671} (\bibinfo {year} {2013})}\BibitemShut
  {NoStop}%
\bibitem [{\citenamefont {Butler}\ \emph {et~al.}(2015)\citenamefont {Butler},
  \citenamefont {Frost},\ and\ \citenamefont {Walsh}}]{Butler2015}%
  \BibitemOpen
  \bibfield  {author} {\bibinfo {author} {\bibfnamefont {K.~T.}\ \bibnamefont
  {Butler}}, \bibinfo {author} {\bibfnamefont {J.~M.}\ \bibnamefont {Frost}}, \
  and\ \bibinfo {author} {\bibfnamefont {A.}~\bibnamefont {Walsh}},\ }\href
  {\doibase 10.1039/c4ee03523b} {\bibfield  {journal} {\bibinfo  {journal}
  {Energy and Environmental Science}\ }\textbf {\bibinfo {volume} {8}},\
  \bibinfo {pages} {838} (\bibinfo {year} {2015})}\BibitemShut {NoStop}%
\bibitem [{\citenamefont {Qin}\ \emph {et~al.}(2008)\citenamefont {Qin},
  \citenamefont {Yao},\ and\ \citenamefont {Liang}}]{Qin2008}%
  \BibitemOpen
  \bibfield  {author} {\bibinfo {author} {\bibfnamefont {M.}~\bibnamefont
  {Qin}}, \bibinfo {author} {\bibfnamefont {K.}~\bibnamefont {Yao}}, \ and\
  \bibinfo {author} {\bibfnamefont {Y.~C.}\ \bibnamefont {Liang}},\ }\href
  {\doibase 10.1063/1.2990754} {\bibfield  {journal} {\bibinfo  {journal}
  {Applied Physics Letters}\ }\textbf {\bibinfo {volume} {93}},\ \bibinfo
  {pages} {122904} (\bibinfo {year} {2008})}\BibitemShut {NoStop}%
\bibitem [{\citenamefont {Li}\ \emph {et~al.}(2014)\citenamefont {Li},
  \citenamefont {Salvador},\ and\ \citenamefont {Rohrer}}]{Li2014}%
  \BibitemOpen
  \bibfield  {author} {\bibinfo {author} {\bibfnamefont {L.}~\bibnamefont
  {Li}}, \bibinfo {author} {\bibfnamefont {P.~A.}\ \bibnamefont {Salvador}}, \
  and\ \bibinfo {author} {\bibfnamefont {G.~S.}\ \bibnamefont {Rohrer}},\
  }\href {\doibase 10.1039/c3nr03998f} {\bibfield  {journal} {\bibinfo
  {journal} {Nanoscale}\ }\textbf {\bibinfo {volume} {6}},\ \bibinfo {pages}
  {24} (\bibinfo {year} {2014})}\BibitemShut {NoStop}%
\bibitem [{\citenamefont {Scott}(2007)}]{Scott2007}%
  \BibitemOpen
  \bibfield  {author} {\bibinfo {author} {\bibfnamefont {J.~F.}\ \bibnamefont
  {Scott}},\ }\href {\doibase 10.1126/science.1129564} {\bibfield  {journal}
  {\bibinfo  {journal} {Science}\ }\textbf {\bibinfo {volume} {315}},\ \bibinfo
  {pages} {954} (\bibinfo {year} {2007})}\BibitemShut {NoStop}%
\bibitem [{\citenamefont {Sando}\ \emph {et~al.}(2018)\citenamefont {Sando},
  \citenamefont {Yang}, \citenamefont {Paillard}, \citenamefont {Dkhil},
  \citenamefont {Bellaiche},\ and\ \citenamefont {Nagarajan}}]{Sando2018}%
  \BibitemOpen
  \bibfield  {author} {\bibinfo {author} {\bibfnamefont {D.}~\bibnamefont
  {Sando}}, \bibinfo {author} {\bibfnamefont {Y.}~\bibnamefont {Yang}},
  \bibinfo {author} {\bibfnamefont {C.}~\bibnamefont {Paillard}}, \bibinfo
  {author} {\bibfnamefont {B.}~\bibnamefont {Dkhil}}, \bibinfo {author}
  {\bibfnamefont {L.}~\bibnamefont {Bellaiche}}, \ and\ \bibinfo {author}
  {\bibfnamefont {V.}~\bibnamefont {Nagarajan}},\ }\href
  {http://dx.doi.org/10.1063/1.5046559} {\bibfield  {journal} {\bibinfo
  {journal} {Applied Physics Reviews}\ }\textbf {\bibinfo {volume} {5}},\
  \bibinfo {pages} {041108} (\bibinfo {year} {2018})}\BibitemShut {NoStop}%
\bibitem [{\citenamefont {Hu}\ \emph {et~al.}(2009)\citenamefont {Hu},
  \citenamefont {Tian}, \citenamefont {Nysten},\ and\ \citenamefont
  {Jonas}}]{Hu2009}%
  \BibitemOpen
  \bibfield  {author} {\bibinfo {author} {\bibfnamefont {Z.}~\bibnamefont
  {Hu}}, \bibinfo {author} {\bibfnamefont {M.}~\bibnamefont {Tian}}, \bibinfo
  {author} {\bibfnamefont {B.}~\bibnamefont {Nysten}}, \ and\ \bibinfo {author}
  {\bibfnamefont {A.~M.}\ \bibnamefont {Jonas}},\ }\href {\doibase
  10.1038/nmat2339} {\bibfield  {journal} {\bibinfo  {journal} {Nature
  Materials}\ }\textbf {\bibinfo {volume} {8}},\ \bibinfo {pages} {62}
  (\bibinfo {year} {2009})}\BibitemShut {NoStop}%
\bibitem [{\citenamefont {Han}\ \emph {et~al.}(2022)\citenamefont {Han},
  \citenamefont {Ji},\ and\ \citenamefont {Yang}}]{Han2022}%
  \BibitemOpen
  \bibfield  {author} {\bibinfo {author} {\bibfnamefont {X.}~\bibnamefont
  {Han}}, \bibinfo {author} {\bibfnamefont {Y.}~\bibnamefont {Ji}}, \ and\
  \bibinfo {author} {\bibfnamefont {Y.}~\bibnamefont {Yang}},\ }\href {\doibase
  10.1002/adfm.202109625} {\bibfield  {journal} {\bibinfo  {journal} {Advanced
  Functional Materials}\ }\textbf {\bibinfo {volume} {32}},\ \bibinfo {pages}
  {2109625} (\bibinfo {year} {2022})}\BibitemShut {NoStop}%
\bibitem [{\citenamefont {Tan}\ \emph {et~al.}(2022)\citenamefont {Tan},
  \citenamefont {Reyes}, \citenamefont {Men{\'{e}}ndez-Proupin}, \citenamefont
  {Reyes-Lillo}, \citenamefont {Li},\ and\ \citenamefont {Zhang}}]{Tan2022}%
  \BibitemOpen
  \bibfield  {author} {\bibinfo {author} {\bibfnamefont {B.}~\bibnamefont
  {Tan}}, \bibinfo {author} {\bibfnamefont {A.~M.}\ \bibnamefont {Reyes}},
  \bibinfo {author} {\bibfnamefont {E.}~\bibnamefont {Men{\'{e}}ndez-Proupin}},
  \bibinfo {author} {\bibfnamefont {S.~E.}\ \bibnamefont {Reyes-Lillo}},
  \bibinfo {author} {\bibfnamefont {Y.}~\bibnamefont {Li}}, \ and\ \bibinfo
  {author} {\bibfnamefont {Z.}~\bibnamefont {Zhang}},\ }\href {\doibase
  10.1021/acsenergylett.2c01750} {\bibfield  {journal} {\bibinfo  {journal}
  {ACS Energy Letters}\ }\textbf {\bibinfo {volume} {7}},\ \bibinfo {pages}
  {3492} (\bibinfo {year} {2022})}\BibitemShut {NoStop}%
\bibitem [{\citenamefont {Zheng}\ \emph {et~al.}(2015)\citenamefont {Zheng},
  \citenamefont {Tan}, \citenamefont {Liu},\ and\ \citenamefont
  {Rappe}}]{Zheng2015}%
  \BibitemOpen
  \bibfield  {author} {\bibinfo {author} {\bibfnamefont {F.}~\bibnamefont
  {Zheng}}, \bibinfo {author} {\bibfnamefont {L.~Z.}\ \bibnamefont {Tan}},
  \bibinfo {author} {\bibfnamefont {S.}~\bibnamefont {Liu}}, \ and\ \bibinfo
  {author} {\bibfnamefont {A.~M.}\ \bibnamefont {Rappe}},\ }\href {\doibase
  10.1021/acs.nanolett.5b01854} {\bibfield  {journal} {\bibinfo  {journal}
  {Nano Letters}\ }\textbf {\bibinfo {volume} {15}},\ \bibinfo {pages} {7794}
  (\bibinfo {year} {2015})}\BibitemShut {NoStop}%
\bibitem [{\citenamefont {Wessels}(2007)}]{Wessels2007}%
  \BibitemOpen
  \bibfield  {author} {\bibinfo {author} {\bibfnamefont {B.~W.}\ \bibnamefont
  {Wessels}},\ }\href {\doibase 10.1146/annurev.matsci.37.052506.084226}
  {\bibfield  {journal} {\bibinfo  {journal} {Annual Review of Materials
  Research}\ }\textbf {\bibinfo {volume} {37}},\ \bibinfo {pages} {659}
  (\bibinfo {year} {2007})}\BibitemShut {NoStop}%
\bibitem [{\citenamefont {Young}\ and\ \citenamefont
  {Rappe}(2012)}]{Young2012}%
  \BibitemOpen
  \bibfield  {author} {\bibinfo {author} {\bibfnamefont {S.~M.}\ \bibnamefont
  {Young}}\ and\ \bibinfo {author} {\bibfnamefont {A.~M.}\ \bibnamefont
  {Rappe}},\ }\href {\doibase 10.1103/PhysRevLett.109.116601} {\bibfield
  {journal} {\bibinfo  {journal} {Physical Review Letters}\ }\textbf {\bibinfo
  {volume} {109}},\ \bibinfo {pages} {116601} (\bibinfo {year}
  {2012})}\BibitemShut {NoStop}%
\bibitem [{\citenamefont {Dai}\ and\ \citenamefont {Rappe}(2023)}]{Dai2023}%
  \BibitemOpen
  \bibfield  {author} {\bibinfo {author} {\bibfnamefont {Z.}~\bibnamefont
  {Dai}}\ and\ \bibinfo {author} {\bibfnamefont {A.~M.}\ \bibnamefont
  {Rappe}},\ }\href@noop {} {\bibfield  {journal} {\bibinfo  {journal}
  {Chemical Physics Reviews}\ }\textbf {\bibinfo {volume} {4}},\ \bibinfo
  {pages} {011303} (\bibinfo {year} {2023})}\BibitemShut {NoStop}%
\bibitem [{\citenamefont {Daranciang}\ \emph {et~al.}(2012)\citenamefont
  {Daranciang}, \citenamefont {Highland}, \citenamefont {Wen}, \citenamefont
  {Young}, \citenamefont {Brandt}, \citenamefont {Hwang}, \citenamefont
  {Vattilana}, \citenamefont {Nicoul}, \citenamefont {Quirin}, \citenamefont
  {Goodfellow}, \citenamefont {Qi}, \citenamefont {Grinberg}, \citenamefont
  {Fritz}, \citenamefont {Cammarata}, \citenamefont {Zhu}, \citenamefont
  {Lemke}, \citenamefont {Walko}, \citenamefont {Dufresne}, \citenamefont
  {Reis}, \citenamefont {Sokolowski-tinten}, \citenamefont {Nelson},
  \citenamefont {Rappe}, \citenamefont {Li}, \citenamefont {Fuoss},
  \citenamefont {Stephenson},\ and\ \citenamefont
  {Lindenberg}}]{Daranciang2012}%
  \BibitemOpen
  \bibfield  {author} {\bibinfo {author} {\bibfnamefont {D.}~\bibnamefont
  {Daranciang}}, \bibinfo {author} {\bibfnamefont {M.~J.}\ \bibnamefont
  {Highland}}, \bibinfo {author} {\bibfnamefont {H.}~\bibnamefont {Wen}},
  \bibinfo {author} {\bibfnamefont {S.~M.}\ \bibnamefont {Young}}, \bibinfo
  {author} {\bibfnamefont {N.~C.}\ \bibnamefont {Brandt}}, \bibinfo {author}
  {\bibfnamefont {H.~Y.}\ \bibnamefont {Hwang}}, \bibinfo {author}
  {\bibfnamefont {M.}~\bibnamefont {Vattilana}}, \bibinfo {author}
  {\bibfnamefont {M.}~\bibnamefont {Nicoul}}, \bibinfo {author} {\bibfnamefont
  {F.}~\bibnamefont {Quirin}}, \bibinfo {author} {\bibfnamefont
  {J.}~\bibnamefont {Goodfellow}}, \bibinfo {author} {\bibfnamefont
  {T.}~\bibnamefont {Qi}}, \bibinfo {author} {\bibfnamefont {I.}~\bibnamefont
  {Grinberg}}, \bibinfo {author} {\bibfnamefont {D.~M.}\ \bibnamefont {Fritz}},
  \bibinfo {author} {\bibfnamefont {M.}~\bibnamefont {Cammarata}}, \bibinfo
  {author} {\bibfnamefont {D.}~\bibnamefont {Zhu}}, \bibinfo {author}
  {\bibfnamefont {H.~T.}\ \bibnamefont {Lemke}}, \bibinfo {author}
  {\bibfnamefont {D.~A.}\ \bibnamefont {Walko}}, \bibinfo {author}
  {\bibfnamefont {E.~M.}\ \bibnamefont {Dufresne}}, \bibinfo {author}
  {\bibfnamefont {D.~A.}\ \bibnamefont {Reis}}, \bibinfo {author}
  {\bibfnamefont {K.}~\bibnamefont {Sokolowski-tinten}}, \bibinfo {author}
  {\bibfnamefont {K.~A.}\ \bibnamefont {Nelson}}, \bibinfo {author}
  {\bibfnamefont {A.~M.}\ \bibnamefont {Rappe}}, \bibinfo {author}
  {\bibfnamefont {Y.}~\bibnamefont {Li}}, \bibinfo {author} {\bibfnamefont
  {P.~H.}\ \bibnamefont {Fuoss}}, \bibinfo {author} {\bibfnamefont {G.~B.}\
  \bibnamefont {Stephenson}}, \ and\ \bibinfo {author} {\bibfnamefont {A.~M.}\
  \bibnamefont {Lindenberg}},\ }\href {\doibase 10.1103/PhysRevLett.108.087601}
  {\bibfield  {journal} {\bibinfo  {journal} {Physical Review Letters}\
  }\textbf {\bibinfo {volume} {108}},\ \bibinfo {pages} {087601} (\bibinfo
  {year} {2012})}\BibitemShut {NoStop}%
\bibitem [{\citenamefont {Kreisel}\ \emph {et~al.}(2012)\citenamefont
  {Kreisel}, \citenamefont {Alexe},\ and\ \citenamefont
  {Thomas}}]{Kreisel2012}%
  \BibitemOpen
  \bibfield  {author} {\bibinfo {author} {\bibfnamefont {J.}~\bibnamefont
  {Kreisel}}, \bibinfo {author} {\bibfnamefont {M.}~\bibnamefont {Alexe}}, \
  and\ \bibinfo {author} {\bibfnamefont {P.~A.}\ \bibnamefont {Thomas}},\
  }\href {\doibase 10.1038/nmat3282} {\bibfield  {journal} {\bibinfo  {journal}
  {Nature Materials}\ }\textbf {\bibinfo {volume} {11}},\ \bibinfo {pages}
  {260} (\bibinfo {year} {2012})}\BibitemShut {NoStop}%
\bibitem [{\citenamefont {Seidel}\ and\ \citenamefont
  {Eng}(2014)}]{Seidel2014}%
  \BibitemOpen
  \bibfield  {author} {\bibinfo {author} {\bibfnamefont {J.}~\bibnamefont
  {Seidel}}\ and\ \bibinfo {author} {\bibfnamefont {L.~M.}\ \bibnamefont
  {Eng}},\ }\href {\doibase 10.1016/j.cap.2014.06.003} {\bibfield  {journal}
  {\bibinfo  {journal} {Current Applied Physics}\ }\textbf {\bibinfo {volume}
  {14}},\ \bibinfo {pages} {1083} (\bibinfo {year} {2014})}\BibitemShut
  {NoStop}%
\bibitem [{\citenamefont {Kim}\ \emph {et~al.}(2018)\citenamefont {Kim},
  \citenamefont {Nguyen},\ and\ \citenamefont {Bark}}]{Kim2018}%
  \BibitemOpen
  \bibfield  {author} {\bibinfo {author} {\bibfnamefont {S.}~\bibnamefont
  {Kim}}, \bibinfo {author} {\bibfnamefont {N.~T.}\ \bibnamefont {Nguyen}}, \
  and\ \bibinfo {author} {\bibfnamefont {C.~W.}\ \bibnamefont {Bark}},\
  }\href@noop {} {\bibfield  {journal} {\bibinfo  {journal} {Applied Sciences}\
  }\textbf {\bibinfo {volume} {8}},\ \bibinfo {pages} {1526} (\bibinfo {year}
  {2018})}\BibitemShut {NoStop}%
\bibitem [{\citenamefont {Li}\ \emph {et~al.}(2017)\citenamefont {Li},
  \citenamefont {Tang}, \citenamefont {Liao}, \citenamefont {Ye}, \citenamefont
  {Zhang}, \citenamefont {Fu}, \citenamefont {You},\ and\ \citenamefont
  {Xiong}}]{Li2017c}%
  \BibitemOpen
  \bibfield  {author} {\bibinfo {author} {\bibfnamefont {P.~F.}\ \bibnamefont
  {Li}}, \bibinfo {author} {\bibfnamefont {Y.~Y.}\ \bibnamefont {Tang}},
  \bibinfo {author} {\bibfnamefont {W.~Q.}\ \bibnamefont {Liao}}, \bibinfo
  {author} {\bibfnamefont {H.~Y.}\ \bibnamefont {Ye}}, \bibinfo {author}
  {\bibfnamefont {Y.}~\bibnamefont {Zhang}}, \bibinfo {author} {\bibfnamefont
  {D.~W.}\ \bibnamefont {Fu}}, \bibinfo {author} {\bibfnamefont {Y.~M.}\
  \bibnamefont {You}}, \ and\ \bibinfo {author} {\bibfnamefont {R.~G.}\
  \bibnamefont {Xiong}},\ }\href {\doibase 10.1038/am.2016.193} {\bibfield
  {journal} {\bibinfo  {journal} {NPG Asia Materials}\ }\textbf {\bibinfo
  {volume} {9}},\ \bibinfo {pages} {1} (\bibinfo {year} {2017})}\BibitemShut
  {NoStop}%
\bibitem [{\citenamefont {Choi}\ \emph
  {et~al.}(2012{\natexlab{a}})\citenamefont {Choi}, \citenamefont {Lee},\ and\
  \citenamefont {Ti}}]{Choi2012}%
  \BibitemOpen
  \bibfield  {author} {\bibinfo {author} {\bibfnamefont {W.~S.}\ \bibnamefont
  {Choi}}, \bibinfo {author} {\bibfnamefont {H.~N.}\ \bibnamefont {Lee}}, \
  and\ \bibinfo {author} {\bibfnamefont {T.~M.}\ \bibnamefont {Ti}},\ }\href
  {\doibase 10.1063/1.3697645} {\bibfield  {journal} {\bibinfo  {journal}
  {Applied Physics Letters}\ }\textbf {\bibinfo {volume} {3}},\ \bibinfo
  {pages} {100} (\bibinfo {year} {2012}{\natexlab{a}})}\BibitemShut {NoStop}%
\bibitem [{\citenamefont {Bennett}\ and\ \citenamefont
  {Rabe}(2012)}]{Bennett2012}%
  \BibitemOpen
  \bibfield  {author} {\bibinfo {author} {\bibfnamefont {J.~W.}\ \bibnamefont
  {Bennett}}\ and\ \bibinfo {author} {\bibfnamefont {K.~M.}\ \bibnamefont
  {Rabe}},\ }\href {\doibase 10.1016/j.jssc.2012.05.013} {\bibfield  {journal}
  {\bibinfo  {journal} {Journal of Solid State Chemistry}\ }\textbf {\bibinfo
  {volume} {195}},\ \bibinfo {pages} {21} (\bibinfo {year} {2012})}\BibitemShut
  {NoStop}%
\bibitem [{\citenamefont {Guo}\ \emph {et~al.}(2013)\citenamefont {Guo},
  \citenamefont {You}, \citenamefont {Zhou}, \citenamefont {Lim}, \citenamefont
  {Zou}, \citenamefont {Chen}, \citenamefont {Ramesh},\ and\ \citenamefont
  {Wang}}]{Guo2013a}%
  \BibitemOpen
  \bibfield  {author} {\bibinfo {author} {\bibfnamefont {R.}~\bibnamefont
  {Guo}}, \bibinfo {author} {\bibfnamefont {L.}~\bibnamefont {You}}, \bibinfo
  {author} {\bibfnamefont {Y.}~\bibnamefont {Zhou}}, \bibinfo {author}
  {\bibfnamefont {Z.~S.}\ \bibnamefont {Lim}}, \bibinfo {author} {\bibfnamefont
  {X.}~\bibnamefont {Zou}}, \bibinfo {author} {\bibfnamefont {L.}~\bibnamefont
  {Chen}}, \bibinfo {author} {\bibfnamefont {R.}~\bibnamefont {Ramesh}}, \ and\
  \bibinfo {author} {\bibfnamefont {J.}~\bibnamefont {Wang}},\ }\href {\doibase
  10.1038/ncomms2990} {\bibfield  {journal} {\bibinfo  {journal} {Nature
  Communications}\ }\textbf {\bibinfo {volume} {4}},\ \bibinfo {pages} {1990}
  (\bibinfo {year} {2013})}\BibitemShut {NoStop}%
\bibitem [{\citenamefont {Grinberg}(2020)}]{Grinberg2020}%
  \BibitemOpen
  \bibfield  {author} {\bibinfo {author} {\bibfnamefont {I.}~\bibnamefont
  {Grinberg}},\ }\href {\doibase 10.1002/ijch.201900124} {\bibfield  {journal}
  {\bibinfo  {journal} {Israel Journal of Chemistry}\ }\textbf {\bibinfo
  {volume} {60}},\ \bibinfo {pages} {823} (\bibinfo {year} {2020})}\BibitemShut
  {NoStop}%
\bibitem [{\citenamefont {Yang}\ \emph {et~al.}(2012)\citenamefont {Yang},
  \citenamefont {Su}, \citenamefont {Shen}, \citenamefont {Zheng},
  \citenamefont {Xin}, \citenamefont {Zhang}, \citenamefont {Hua},
  \citenamefont {Chen},\ and\ \citenamefont {Harris}}]{Yang2012}%
  \BibitemOpen
  \bibfield  {author} {\bibinfo {author} {\bibfnamefont {X.}~\bibnamefont
  {Yang}}, \bibinfo {author} {\bibfnamefont {X.}~\bibnamefont {Su}}, \bibinfo
  {author} {\bibfnamefont {M.}~\bibnamefont {Shen}}, \bibinfo {author}
  {\bibfnamefont {F.}~\bibnamefont {Zheng}}, \bibinfo {author} {\bibfnamefont
  {Y.}~\bibnamefont {Xin}}, \bibinfo {author} {\bibfnamefont {L.}~\bibnamefont
  {Zhang}}, \bibinfo {author} {\bibfnamefont {M.}~\bibnamefont {Hua}}, \bibinfo
  {author} {\bibfnamefont {Y.}~\bibnamefont {Chen}}, \ and\ \bibinfo {author}
  {\bibfnamefont {V.~G.}\ \bibnamefont {Harris}},\ }\href {\doibase
  10.1002/adma.201104078} {\bibfield  {journal} {\bibinfo  {journal} {Advanced
  Materials}\ }\textbf {\bibinfo {volume} {24}},\ \bibinfo {pages} {1202}
  (\bibinfo {year} {2012})}\BibitemShut {NoStop}%
\bibitem [{\citenamefont {Bennett}\ \emph {et~al.}(2012)\citenamefont
  {Bennett}, \citenamefont {Garrity}, \citenamefont {Rabe},\ and\ \citenamefont
  {Vanderbilt}}]{Bennett2012d}%
  \BibitemOpen
  \bibfield  {author} {\bibinfo {author} {\bibfnamefont {J.~W.}\ \bibnamefont
  {Bennett}}, \bibinfo {author} {\bibfnamefont {K.~F.}\ \bibnamefont
  {Garrity}}, \bibinfo {author} {\bibfnamefont {K.~M.}\ \bibnamefont {Rabe}}, \
  and\ \bibinfo {author} {\bibfnamefont {D.}~\bibnamefont {Vanderbilt}},\
  }\href {\doibase 10.1103/PhysRevLett.109.167602} {\bibfield  {journal}
  {\bibinfo  {journal} {Physical Review Letters}\ }\textbf {\bibinfo {volume}
  {109}},\ \bibinfo {pages} {167602} (\bibinfo {year} {2012})}\BibitemShut
  {NoStop}%
\bibitem [{\citenamefont {Wang}\ \emph {et~al.}(2015)\citenamefont {Wang},
  \citenamefont {Grinberg}, \citenamefont {Jiang}, \citenamefont {Young},
  \citenamefont {Davies},\ and\ \citenamefont {Rappe}}]{Wang2015b}%
  \BibitemOpen
  \bibfield  {author} {\bibinfo {author} {\bibfnamefont {F.}~\bibnamefont
  {Wang}}, \bibinfo {author} {\bibfnamefont {I.}~\bibnamefont {Grinberg}},
  \bibinfo {author} {\bibfnamefont {L.}~\bibnamefont {Jiang}}, \bibinfo
  {author} {\bibfnamefont {S.~M.}\ \bibnamefont {Young}}, \bibinfo {author}
  {\bibfnamefont {P.~K.}\ \bibnamefont {Davies}}, \ and\ \bibinfo {author}
  {\bibfnamefont {A.~M.}\ \bibnamefont {Rappe}},\ }\href {\doibase
  10.1080/00150193.2015.1058096} {\bibfield  {journal} {\bibinfo  {journal}
  {Ferroelectrics}\ }\textbf {\bibinfo {volume} {483}},\ \bibinfo {pages} {1}
  (\bibinfo {year} {2015})}\BibitemShut {NoStop}%
\bibitem [{\citenamefont {Zhang}\ \emph {et~al.}(2017)\citenamefont {Zhang},
  \citenamefont {Sahoo},\ and\ \citenamefont {Wang}}]{Zhang2017}%
  \BibitemOpen
  \bibfield  {author} {\bibinfo {author} {\bibfnamefont {Y.}~\bibnamefont
  {Zhang}}, \bibinfo {author} {\bibfnamefont {M.~P.}\ \bibnamefont {Sahoo}}, \
  and\ \bibinfo {author} {\bibfnamefont {J.}~\bibnamefont {Wang}},\ }\href
  {\doibase 10.1039/c6cp06042k} {\bibfield  {journal} {\bibinfo  {journal}
  {Physical Chemistry Chemical Physics}\ }\textbf {\bibinfo {volume} {19}},\
  \bibinfo {pages} {7032} (\bibinfo {year} {2017})}\BibitemShut {NoStop}%
\bibitem [{\citenamefont {Wang}\ \emph {et~al.}(2014)\citenamefont {Wang},
  \citenamefont {Grinberg},\ and\ \citenamefont {Rappe}}]{Wang2014a}%
  \BibitemOpen
  \bibfield  {author} {\bibinfo {author} {\bibfnamefont {F.}~\bibnamefont
  {Wang}}, \bibinfo {author} {\bibfnamefont {I.}~\bibnamefont {Grinberg}}, \
  and\ \bibinfo {author} {\bibfnamefont {A.~M.}\ \bibnamefont {Rappe}},\ }\href
  {\doibase 10.1063/1.4871707} {\bibfield  {journal} {\bibinfo  {journal}
  {Applied Physics Letters}\ }\textbf {\bibinfo {volume} {104}},\ \bibinfo
  {pages} {152903} (\bibinfo {year} {2014})}\BibitemShut {NoStop}%
\bibitem [{\citenamefont {Ghosh}\ \emph {et~al.}(2021)\citenamefont {Ghosh},
  \citenamefont {Doherty}, \citenamefont {Lisenkov},\ and\ \citenamefont
  {Ponomareva}}]{Ghosh2021}%
  \BibitemOpen
  \bibfield  {author} {\bibinfo {author} {\bibfnamefont {P.~S.}\ \bibnamefont
  {Ghosh}}, \bibinfo {author} {\bibfnamefont {J.}~\bibnamefont {Doherty}},
  \bibinfo {author} {\bibfnamefont {S.}~\bibnamefont {Lisenkov}}, \ and\
  \bibinfo {author} {\bibfnamefont {I.}~\bibnamefont {Ponomareva}},\ }\href
  {\doibase 10.1021/acs.jpcc.1c03980} {\bibfield  {journal} {\bibinfo
  {journal} {Journal of Physical Chemistry C}\ }\textbf {\bibinfo {volume}
  {125}},\ \bibinfo {pages} {16296} (\bibinfo {year} {2021})}\BibitemShut
  {NoStop}%
\bibitem [{\citenamefont {Bonomi}\ \emph {et~al.}(2018)\citenamefont {Bonomi},
  \citenamefont {Tredici}, \citenamefont {Albini}, \citenamefont {Galinetto},
  \citenamefont {Rizzo}, \citenamefont {Listorti}, \citenamefont {Tamburini},\
  and\ \citenamefont {Malavasi}}]{Bonomi2018}%
  \BibitemOpen
  \bibfield  {author} {\bibinfo {author} {\bibfnamefont {S.}~\bibnamefont
  {Bonomi}}, \bibinfo {author} {\bibfnamefont {I.}~\bibnamefont {Tredici}},
  \bibinfo {author} {\bibfnamefont {B.}~\bibnamefont {Albini}}, \bibinfo
  {author} {\bibfnamefont {P.}~\bibnamefont {Galinetto}}, \bibinfo {author}
  {\bibfnamefont {A.}~\bibnamefont {Rizzo}}, \bibinfo {author} {\bibfnamefont
  {A.}~\bibnamefont {Listorti}}, \bibinfo {author} {\bibfnamefont
  {A.}~\bibnamefont {Tamburini}}, \ and\ \bibinfo {author} {\bibfnamefont
  {L.}~\bibnamefont {Malavasi}},\ }\href {\doibase 10.1039/c8cc08549h}
  {\bibfield  {journal} {\bibinfo  {journal} {ChemComm}\ }\textbf {\bibinfo
  {volume} {54}},\ \bibinfo {pages} {13212} (\bibinfo {year}
  {2018})}\BibitemShut {NoStop}%
\bibitem [{\citenamefont {Yang}\ \emph {et~al.}(2016)\citenamefont {Yang},
  \citenamefont {Wang}, \citenamefont {Yan},\ and\ \citenamefont
  {Chen}}]{Yang2016}%
  \BibitemOpen
  \bibfield  {author} {\bibinfo {author} {\bibfnamefont {X.}~\bibnamefont
  {Yang}}, \bibinfo {author} {\bibfnamefont {Y.}~\bibnamefont {Wang}}, \bibinfo
  {author} {\bibfnamefont {H.}~\bibnamefont {Yan}}, \ and\ \bibinfo {author}
  {\bibfnamefont {Y.}~\bibnamefont {Chen}},\ }\href {\doibase
  10.1016/j.commatsci.2016.04.021} {\bibfield  {journal} {\bibinfo  {journal}
  {Computational Materials Science}\ }\textbf {\bibinfo {volume} {121}},\
  \bibinfo {pages} {61} (\bibinfo {year} {2016})}\BibitemShut {NoStop}%
\bibitem [{\citenamefont {Berger}\ \emph {et~al.}(2011)\citenamefont {Berger},
  \citenamefont {Fennie},\ and\ \citenamefont {Neaton}}]{Berger2011}%
  \BibitemOpen
  \bibfield  {author} {\bibinfo {author} {\bibfnamefont {R.~F.}\ \bibnamefont
  {Berger}}, \bibinfo {author} {\bibfnamefont {C.~J.}\ \bibnamefont {Fennie}},
  \ and\ \bibinfo {author} {\bibfnamefont {J.~B.}\ \bibnamefont {Neaton}},\
  }\href {\doibase 10.1103/PhysRevLett.107.146804} {\bibfield  {journal}
  {\bibinfo  {journal} {Physical Review Letters}\ }\textbf {\bibinfo {volume}
  {107}},\ \bibinfo {pages} {146804} (\bibinfo {year} {2011})}\BibitemShut
  {NoStop}%
\bibitem [{\citenamefont {Vonr{\"{u}}ti}\ and\ \citenamefont
  {Aschauer}(2018)}]{Vonruti2018}%
  \BibitemOpen
  \bibfield  {author} {\bibinfo {author} {\bibfnamefont {N.}~\bibnamefont
  {Vonr{\"{u}}ti}}\ and\ \bibinfo {author} {\bibfnamefont {U.}~\bibnamefont
  {Aschauer}},\ }\href {\doibase 10.1103/PhysRevMaterials.2.105401} {\bibfield
  {journal} {\bibinfo  {journal} {Physical Review Materials}\ }\textbf
  {\bibinfo {volume} {2}},\ \bibinfo {pages} {105401} (\bibinfo {year}
  {2018})}\BibitemShut {NoStop}%
\bibitem [{\citenamefont {Manzeli}\ \emph {et~al.}(2015)\citenamefont
  {Manzeli}, \citenamefont {Allain}, \citenamefont {Ghadimi},\ and\
  \citenamefont {Kis}}]{Manzeli2015}%
  \BibitemOpen
  \bibfield  {author} {\bibinfo {author} {\bibfnamefont {S.}~\bibnamefont
  {Manzeli}}, \bibinfo {author} {\bibfnamefont {A.}~\bibnamefont {Allain}},
  \bibinfo {author} {\bibfnamefont {A.}~\bibnamefont {Ghadimi}}, \ and\
  \bibinfo {author} {\bibfnamefont {A.}~\bibnamefont {Kis}},\ }\href {\doibase
  10.1021/acs.nanolett.5b01689} {\bibfield  {journal} {\bibinfo  {journal}
  {Nano Letters}\ }\textbf {\bibinfo {volume} {15}},\ \bibinfo {pages} {5330}
  (\bibinfo {year} {2015})}\BibitemShut {NoStop}%
\bibitem [{\citenamefont {Reyes-Lillo}\ \emph {et~al.}(2016)\citenamefont
  {Reyes-Lillo}, \citenamefont {Rangel}, \citenamefont {Bruneval},\ and\
  \citenamefont {Neaton}}]{Reyes-Lillo2016}%
  \BibitemOpen
  \bibfield  {author} {\bibinfo {author} {\bibfnamefont {S.~E.}\ \bibnamefont
  {Reyes-Lillo}}, \bibinfo {author} {\bibfnamefont {T.}~\bibnamefont {Rangel}},
  \bibinfo {author} {\bibfnamefont {F.}~\bibnamefont {Bruneval}}, \ and\
  \bibinfo {author} {\bibfnamefont {J.~B.}\ \bibnamefont {Neaton}},\ }\href
  {\doibase 10.1103/PhysRevB.94.041107} {\bibfield  {journal} {\bibinfo
  {journal} {Physical Review B}\ }\textbf {\bibinfo {volume} {94}},\ \bibinfo
  {pages} {041107} (\bibinfo {year} {2016})}\BibitemShut {NoStop}%
\bibitem [{\citenamefont {Birol}\ and\ \citenamefont
  {Fennie}(2013)}]{Birol2013}%
  \BibitemOpen
  \bibfield  {author} {\bibinfo {author} {\bibfnamefont {T.}~\bibnamefont
  {Birol}}\ and\ \bibinfo {author} {\bibfnamefont {C.~J.}\ \bibnamefont
  {Fennie}},\ }\href {\doibase 10.1103/PhysRevB.88.094103} {\bibfield
  {journal} {\bibinfo  {journal} {Physical Review B - Condensed Matter and
  Materials Physics}\ }\textbf {\bibinfo {volume} {88}},\ \bibinfo {pages}
  {094103} (\bibinfo {year} {2013})}\BibitemShut {NoStop}%
\bibitem [{\citenamefont {Shimada}\ \emph {et~al.}(2011)\citenamefont
  {Shimada}, \citenamefont {Zenpuku}, \citenamefont {Fujiwara}, \citenamefont
  {Hazu}, \citenamefont {Chichibu}, \citenamefont {Hata}, \citenamefont
  {Sazawa}, \citenamefont {Takada},\ and\ \citenamefont {Sota}}]{Shimada2011}%
  \BibitemOpen
  \bibfield  {author} {\bibinfo {author} {\bibfnamefont {K.}~\bibnamefont
  {Shimada}}, \bibinfo {author} {\bibfnamefont {A.}~\bibnamefont {Zenpuku}},
  \bibinfo {author} {\bibfnamefont {K.}~\bibnamefont {Fujiwara}}, \bibinfo
  {author} {\bibfnamefont {K.}~\bibnamefont {Hazu}}, \bibinfo {author}
  {\bibfnamefont {S.~F.}\ \bibnamefont {Chichibu}}, \bibinfo {author}
  {\bibfnamefont {M.}~\bibnamefont {Hata}}, \bibinfo {author} {\bibfnamefont
  {H.}~\bibnamefont {Sazawa}}, \bibinfo {author} {\bibfnamefont
  {T.}~\bibnamefont {Takada}}, \ and\ \bibinfo {author} {\bibfnamefont
  {T.}~\bibnamefont {Sota}},\ }\href@noop {} {\bibfield  {journal} {\bibinfo
  {journal} {Journal of Applied Physics}\ }\textbf {\bibinfo {volume} {110}},\
  \bibinfo {pages} {074114} (\bibinfo {year} {2011})}\BibitemShut {NoStop}%
\bibitem [{\citenamefont {Qi}\ \emph {et~al.}(2011)\citenamefont {Qi},
  \citenamefont {Grinberg},\ and\ \citenamefont {Rappe}}]{Qi2011}%
  \BibitemOpen
  \bibfield  {author} {\bibinfo {author} {\bibfnamefont {T.}~\bibnamefont
  {Qi}}, \bibinfo {author} {\bibfnamefont {I.}~\bibnamefont {Grinberg}}, \ and\
  \bibinfo {author} {\bibfnamefont {A.~M.}\ \bibnamefont {Rappe}},\ }\href
  {\doibase 10.1103/PhysRevB.83.224108} {\bibfield  {journal} {\bibinfo
  {journal} {Physical Review B - Condensed Matter and Materials Physics}\
  }\textbf {\bibinfo {volume} {83}},\ \bibinfo {pages} {224108} (\bibinfo
  {year} {2011})}\BibitemShut {NoStop}%
\bibitem [{\citenamefont {Islam}\ and\ \citenamefont
  {Podder}(2021)}]{Islam2021}%
  \BibitemOpen
  \bibfield  {author} {\bibinfo {author} {\bibfnamefont {M.~N.}\ \bibnamefont
  {Islam}}\ and\ \bibinfo {author} {\bibfnamefont {J.}~\bibnamefont {Podder}},\
  }\href {\doibase 10.1016/j.mssp.2020.105419} {\bibfield  {journal} {\bibinfo
  {journal} {Materials Science in Semiconductor Processing}\ }\textbf {\bibinfo
  {volume} {121}},\ \bibinfo {pages} {105419} (\bibinfo {year}
  {2021})}\BibitemShut {NoStop}%
\bibitem [{\citenamefont {Wang}\ \emph {et~al.}(2022)\citenamefont {Wang},
  \citenamefont {Bruy{\`{e}}re}, \citenamefont {Kumagai}, \citenamefont
  {Tsunoda}, \citenamefont {Oba}, \citenamefont {Ghanbaja}, \citenamefont
  {Sun}, \citenamefont {Dai},\ and\ \citenamefont {Pierson}}]{Wang2022a}%
  \BibitemOpen
  \bibfield  {author} {\bibinfo {author} {\bibfnamefont {Y.}~\bibnamefont
  {Wang}}, \bibinfo {author} {\bibfnamefont {S.}~\bibnamefont {Bruy{\`{e}}re}},
  \bibinfo {author} {\bibfnamefont {Y.}~\bibnamefont {Kumagai}}, \bibinfo
  {author} {\bibfnamefont {N.}~\bibnamefont {Tsunoda}}, \bibinfo {author}
  {\bibfnamefont {F.}~\bibnamefont {Oba}}, \bibinfo {author} {\bibfnamefont
  {J.}~\bibnamefont {Ghanbaja}}, \bibinfo {author} {\bibfnamefont
  {H.}~\bibnamefont {Sun}}, \bibinfo {author} {\bibfnamefont {B.}~\bibnamefont
  {Dai}}, \ and\ \bibinfo {author} {\bibfnamefont {J.~F.}\ \bibnamefont
  {Pierson}},\ }\href {\doibase 10.1039/d2ra01887j} {\bibfield  {journal}
  {\bibinfo  {journal} {RSC Advances}\ }\textbf {\bibinfo {volume} {12}},\
  \bibinfo {pages} {21940} (\bibinfo {year} {2022})}\BibitemShut {NoStop}%
\bibitem [{\citenamefont {Choi}\ \emph
  {et~al.}(2012{\natexlab{b}})\citenamefont {Choi}, \citenamefont {Chisholm},
  \citenamefont {Singh}, \citenamefont {Choi}, \citenamefont {Jellison},\ and\
  \citenamefont {Lee}}]{Choi2012a}%
  \BibitemOpen
  \bibfield  {author} {\bibinfo {author} {\bibfnamefont {W.~S.}\ \bibnamefont
  {Choi}}, \bibinfo {author} {\bibfnamefont {M.~F.}\ \bibnamefont {Chisholm}},
  \bibinfo {author} {\bibfnamefont {D.~J.}\ \bibnamefont {Singh}}, \bibinfo
  {author} {\bibfnamefont {T.}~\bibnamefont {Choi}}, \bibinfo {author}
  {\bibfnamefont {G.~E.}\ \bibnamefont {Jellison}}, \ and\ \bibinfo {author}
  {\bibfnamefont {H.~N.}\ \bibnamefont {Lee}},\ }\href {\doibase
  10.1038/ncomms1690} {\bibfield  {journal} {\bibinfo  {journal} {Nature
  Communications}\ }\textbf {\bibinfo {volume} {3}},\ \bibinfo {pages} {686}
  (\bibinfo {year} {2012}{\natexlab{b}})}\BibitemShut {NoStop}%
\bibitem [{\citenamefont {He}\ \emph {et~al.}(2017)\citenamefont {He},
  \citenamefont {Franchini},\ and\ \citenamefont {Rondinelli}}]{He2017}%
  \BibitemOpen
  \bibfield  {author} {\bibinfo {author} {\bibfnamefont {J.}~\bibnamefont
  {He}}, \bibinfo {author} {\bibfnamefont {C.}~\bibnamefont {Franchini}}, \
  and\ \bibinfo {author} {\bibfnamefont {J.~M.}\ \bibnamefont {Rondinelli}},\
  }\href {\doibase 10.1021/acs.chemmater.6b03486} {\bibfield  {journal}
  {\bibinfo  {journal} {Chemistry of Materials}\ }\textbf {\bibinfo {volume}
  {29}},\ \bibinfo {pages} {2445} (\bibinfo {year} {2017})}\BibitemShut
  {NoStop}%
\bibitem [{\citenamefont {Parker}\ \emph {et~al.}(2011)\citenamefont {Parker},
  \citenamefont {Rondinelli},\ and\ \citenamefont {Nakhmanson}}]{Parker2011}%
  \BibitemOpen
  \bibfield  {author} {\bibinfo {author} {\bibfnamefont {W.~D.}\ \bibnamefont
  {Parker}}, \bibinfo {author} {\bibfnamefont {J.~M.}\ \bibnamefont
  {Rondinelli}}, \ and\ \bibinfo {author} {\bibfnamefont {S.~M.}\ \bibnamefont
  {Nakhmanson}},\ }\href {\doibase 10.1103/PhysRevB.84.245126} {\bibfield
  {journal} {\bibinfo  {journal} {Physical Review B - Condensed Matter and
  Materials Physics}\ }\textbf {\bibinfo {volume} {84}},\ \bibinfo {pages}
  {245126} (\bibinfo {year} {2011})}\BibitemShut {NoStop}%
\bibitem [{\citenamefont {Fridkin}(1979)}]{Fridkinbook}%
  \BibitemOpen
  \bibfield  {author} {\bibinfo {author} {\bibfnamefont {V.~M.}\ \bibnamefont
  {Fridkin}},\ }\href@noop {} { {\bibinfo {title}
  {{Photoferroelectrics}}}}\ (\bibinfo  {publisher} {Springer Berlin
  Heidelberg},\ \bibinfo {year} {1979})\BibitemShut {NoStop}%
\bibitem [{\citenamefont {Ma}\ \emph {et~al.}(2020)\citenamefont {Ma},
  \citenamefont {Yang}, \citenamefont {Lei}, \citenamefont {Zheng},
  \citenamefont {Chen},\ and\ \citenamefont {Song}}]{Ma2020}%
  \BibitemOpen
  \bibfield  {author} {\bibinfo {author} {\bibfnamefont {X.}~\bibnamefont
  {Ma}}, \bibinfo {author} {\bibfnamefont {L.}~\bibnamefont {Yang}}, \bibinfo
  {author} {\bibfnamefont {K.}~\bibnamefont {Lei}}, \bibinfo {author}
  {\bibfnamefont {S.}~\bibnamefont {Zheng}}, \bibinfo {author} {\bibfnamefont
  {C.}~\bibnamefont {Chen}}, \ and\ \bibinfo {author} {\bibfnamefont
  {H.}~\bibnamefont {Song}},\ }\href {\doibase 10.1016/j.nanoen.2020.105354}
  {\bibfield  {journal} {\bibinfo  {journal} {Nano Energy}\ }\textbf {\bibinfo
  {volume} {78}},\ \bibinfo {pages} {105354} (\bibinfo {year}
  {2020})}\BibitemShut {NoStop}%
\bibitem [{\citenamefont {Rus}\ \emph {et~al.}(2016)\citenamefont {Rus},
  \citenamefont {Ward},\ and\ \citenamefont {Herklotz}}]{Rus2016}%
  \BibitemOpen
  \bibfield  {author} {\bibinfo {author} {\bibfnamefont {S.~F.}\ \bibnamefont
  {Rus}}, \bibinfo {author} {\bibfnamefont {T.~Z.}\ \bibnamefont {Ward}}, \
  and\ \bibinfo {author} {\bibfnamefont {A.}~\bibnamefont {Herklotz}},\ }\href
  {\doibase 10.1016/j.tsf.2016.06.057} {\bibfield  {journal} {\bibinfo
  {journal} {Thin Solid Films}\ }\textbf {\bibinfo {volume} {615}},\ \bibinfo
  {pages} {103} (\bibinfo {year} {2016})}\BibitemShut {NoStop}%
\bibitem [{\citenamefont {Smidt}\ \emph {et~al.}(2020)\citenamefont {Smidt},
  \citenamefont {Mack}, \citenamefont {Reyes-Lillo}, \citenamefont {Jain},\
  and\ \citenamefont {Neaton}}]{Smidt2020}%
  \BibitemOpen
  \bibfield  {author} {\bibinfo {author} {\bibfnamefont {T.~E.}\ \bibnamefont
  {Smidt}}, \bibinfo {author} {\bibfnamefont {S.~A.}\ \bibnamefont {Mack}},
  \bibinfo {author} {\bibfnamefont {S.~E.}\ \bibnamefont {Reyes-Lillo}},
  \bibinfo {author} {\bibfnamefont {A.}~\bibnamefont {Jain}}, \ and\ \bibinfo
  {author} {\bibfnamefont {J.~B.}\ \bibnamefont {Neaton}},\ }\href@noop {}
  {\bibfield  {journal} {\bibinfo  {journal} {Scientific Data}\ }\textbf
  {\bibinfo {volume} {7}},\ \bibinfo {pages} {72} (\bibinfo {year}
  {2020})}\BibitemShut {NoStop}%
\bibitem [{\citenamefont {Ricci}\ \emph {et~al.}(2024)\citenamefont {Ricci},
  \citenamefont {Reyes-Lillo}, \citenamefont {Mack},\ and\ \citenamefont
  {Neaton}}]{Ricci2024}%
  \BibitemOpen
  \bibfield  {author} {\bibinfo {author} {\bibfnamefont {F.}~\bibnamefont
  {Ricci}}, \bibinfo {author} {\bibfnamefont {S.~E.}\ \bibnamefont
  {Reyes-Lillo}}, \bibinfo {author} {\bibfnamefont {S.~A.}\ \bibnamefont
  {Mack}}, \ and\ \bibinfo {author} {\bibfnamefont {J.~B.}\ \bibnamefont
  {Neaton}},\ }\href {\doibase 10.1038/s41524-023-01193-3} {\bibfield
  {journal} {\bibinfo  {journal} {npj Computational Materials}\ }\textbf
  {\bibinfo {volume} {10}},\ \bibinfo {pages} {1} (\bibinfo {year}
  {2024})}\BibitemShut {NoStop}%
\bibitem [{\citenamefont {Jain}\ \emph {et~al.}(2013)\citenamefont {Jain},
  \citenamefont {Ong}, \citenamefont {Hautier}, \citenamefont {Chen},
  \citenamefont {Richards}, \citenamefont {Dacek}, \citenamefont {Cholia},
  \citenamefont {Gunter}, \citenamefont {Skinner}, \citenamefont {Ceder},\ and\
  \citenamefont {Persson}}]{Jain2013}%
  \BibitemOpen
  \bibfield  {author} {\bibinfo {author} {\bibfnamefont {A.}~\bibnamefont
  {Jain}}, \bibinfo {author} {\bibfnamefont {S.~P.}\ \bibnamefont {Ong}},
  \bibinfo {author} {\bibfnamefont {G.}~\bibnamefont {Hautier}}, \bibinfo
  {author} {\bibfnamefont {W.}~\bibnamefont {Chen}}, \bibinfo {author}
  {\bibfnamefont {W.~D.}\ \bibnamefont {Richards}}, \bibinfo {author}
  {\bibfnamefont {S.}~\bibnamefont {Dacek}}, \bibinfo {author} {\bibfnamefont
  {S.}~\bibnamefont {Cholia}}, \bibinfo {author} {\bibfnamefont
  {D.}~\bibnamefont {Gunter}}, \bibinfo {author} {\bibfnamefont
  {D.}~\bibnamefont {Skinner}}, \bibinfo {author} {\bibfnamefont
  {G.}~\bibnamefont {Ceder}}, \ and\ \bibinfo {author} {\bibfnamefont {K.~A.}\
  \bibnamefont {Persson}},\ }\href@noop {} {\bibfield  {journal} {\bibinfo
  {journal} {APL Materials}\ }\textbf {\bibinfo {volume} {1}} (\bibinfo {year}
  {2013})}\BibitemShut {NoStop}%
\bibitem [{\citenamefont {Jain}\ \emph
  {et~al.}(2011{\natexlab{a}})\citenamefont {Jain}, \citenamefont {Hautier},
  \citenamefont {Moore}, \citenamefont {{Ping Ong}}, \citenamefont {Fischer},
  \citenamefont {Mueller}, \citenamefont {Persson},\ and\ \citenamefont
  {Ceder}}]{Jain2011}%
  \BibitemOpen
  \bibfield  {author} {\bibinfo {author} {\bibfnamefont {A.}~\bibnamefont
  {Jain}}, \bibinfo {author} {\bibfnamefont {G.}~\bibnamefont {Hautier}},
  \bibinfo {author} {\bibfnamefont {C.~J.}\ \bibnamefont {Moore}}, \bibinfo
  {author} {\bibfnamefont {S.}~\bibnamefont {{Ping Ong}}}, \bibinfo {author}
  {\bibfnamefont {C.~C.}\ \bibnamefont {Fischer}}, \bibinfo {author}
  {\bibfnamefont {T.}~\bibnamefont {Mueller}}, \bibinfo {author} {\bibfnamefont
  {K.~A.}\ \bibnamefont {Persson}}, \ and\ \bibinfo {author} {\bibfnamefont
  {G.}~\bibnamefont {Ceder}},\ }\href {\doibase
  10.1016/j.commatsci.2011.02.023} {\bibfield  {journal} {\bibinfo  {journal}
  {Computational Materials Science}\ }\textbf {\bibinfo {volume} {50}},\
  \bibinfo {pages} {2295} (\bibinfo {year} {2011}{\natexlab{a}})}\BibitemShut
  {NoStop}%
\bibitem [{\citenamefont {Ong}\ \emph {et~al.}(2013)\citenamefont {Ong},
  \citenamefont {Richards}, \citenamefont {Jain}, \citenamefont {Hautier},
  \citenamefont {Kocher}, \citenamefont {Cholia}, \citenamefont {Gunter},
  \citenamefont {Chevrier}, \citenamefont {Persson},\ and\ \citenamefont
  {Ceder}}]{Ong2013}%
  \BibitemOpen
  \bibfield  {author} {\bibinfo {author} {\bibfnamefont {S.~P.}\ \bibnamefont
  {Ong}}, \bibinfo {author} {\bibfnamefont {W.~D.}\ \bibnamefont {Richards}},
  \bibinfo {author} {\bibfnamefont {A.}~\bibnamefont {Jain}}, \bibinfo {author}
  {\bibfnamefont {G.}~\bibnamefont {Hautier}}, \bibinfo {author} {\bibfnamefont
  {M.}~\bibnamefont {Kocher}}, \bibinfo {author} {\bibfnamefont
  {S.}~\bibnamefont {Cholia}}, \bibinfo {author} {\bibfnamefont
  {D.}~\bibnamefont {Gunter}}, \bibinfo {author} {\bibfnamefont {V.~L.}\
  \bibnamefont {Chevrier}}, \bibinfo {author} {\bibfnamefont {K.~A.}\
  \bibnamefont {Persson}}, \ and\ \bibinfo {author} {\bibfnamefont
  {G.}~\bibnamefont {Ceder}},\ }\href {\doibase
  10.1016/j.commatsci.2012.10.028} {\bibfield  {journal} {\bibinfo  {journal}
  {Computational Materials Science}\ }\textbf {\bibinfo {volume} {68}},\
  \bibinfo {pages} {314} (\bibinfo {year} {2013})}\BibitemShut {NoStop}%
\bibitem [{\citenamefont {Kresse}\ and\ \citenamefont
  {Furthm{\"{u}}ller}(1996)}]{Kresse1996}%
  \BibitemOpen
  \bibfield  {author} {\bibinfo {author} {\bibfnamefont {G.}~\bibnamefont
  {Kresse}}\ and\ \bibinfo {author} {\bibfnamefont {J.}~\bibnamefont
  {Furthm{\"{u}}ller}},\ }\href {\doibase 10.1016/0927-0256(96)00008-0}
  {\bibfield  {journal} {\bibinfo  {journal} {Computational Materials Science}\
  }\textbf {\bibinfo {volume} {6}},\ \bibinfo {pages} {15} (\bibinfo {year}
  {1996})}\BibitemShut {NoStop}%
\bibitem [{\citenamefont {Perdew}\ \emph {et~al.}(1996)\citenamefont {Perdew},
  \citenamefont {Burke},\ and\ \citenamefont {Ernzerhof}}]{Perdew1996}%
  \BibitemOpen
  \bibfield  {author} {\bibinfo {author} {\bibfnamefont {J.~P.}\ \bibnamefont
  {Perdew}}, \bibinfo {author} {\bibfnamefont {K.}~\bibnamefont {Burke}}, \
  and\ \bibinfo {author} {\bibfnamefont {M.}~\bibnamefont {Ernzerhof}},\
  }\href@noop {} {\bibfield  {journal} {\bibinfo  {journal} {Physical Review
  Letters}\ }\textbf {\bibinfo {volume} {18}},\ \bibinfo {pages} {3865}
  (\bibinfo {year} {1996})}\BibitemShut {NoStop}%
\bibitem [{\citenamefont {Liechtenstein}\ \emph {et~al.}(1995)\citenamefont
  {Liechtenstein}, \citenamefont {Anisimov},\ and\ \citenamefont
  {Zaanen}}]{Liechtenstein1995}%
  \BibitemOpen
  \bibfield  {author} {\bibinfo {author} {\bibfnamefont {A.~I.}\ \bibnamefont
  {Liechtenstein}}, \bibinfo {author} {\bibfnamefont {V.}~\bibnamefont
  {Anisimov}}, \ and\ \bibinfo {author} {\bibfnamefont {J.}~\bibnamefont
  {Zaanen}},\ }\href@noop {} {\bibfield  {journal} {\bibinfo  {journal}
  {Physical Review B}\ }\textbf {\bibinfo {volume} {52}},\ \bibinfo {pages}
  {5467} (\bibinfo {year} {1995})}\BibitemShut {NoStop}%
\bibitem [{\citenamefont {Dudarev}\ and\ \citenamefont
  {Botton}(1998)}]{Dudarev1998}%
  \BibitemOpen
  \bibfield  {author} {\bibinfo {author} {\bibfnamefont {S.}~\bibnamefont
  {Dudarev}}\ and\ \bibinfo {author} {\bibfnamefont {G.}~\bibnamefont
  {Botton}},\ }\href {\doibase 10.1103/PhysRevB.57.1505} {\bibfield  {journal}
  {\bibinfo  {journal} {Physical Review B - Condensed Matter and Materials
  Physics}\ }\textbf {\bibinfo {volume} {57}},\ \bibinfo {pages} {1505}
  (\bibinfo {year} {1998})}\BibitemShut {NoStop}%
\bibitem [{\citenamefont {Wang}\ \emph {et~al.}(2006)\citenamefont {Wang},
  \citenamefont {Maxisch},\ and\ \citenamefont {Ceder}}]{Wang2006a}%
  \BibitemOpen
  \bibfield  {author} {\bibinfo {author} {\bibfnamefont {L.}~\bibnamefont
  {Wang}}, \bibinfo {author} {\bibfnamefont {T.}~\bibnamefont {Maxisch}}, \
  and\ \bibinfo {author} {\bibfnamefont {G.}~\bibnamefont {Ceder}},\
  }\href@noop {} {\bibfield  {journal} {\bibinfo  {journal} {Physical Review B
  - Condensed Matter and Materials Physics}\ }\textbf {\bibinfo {volume}
  {73}},\ \bibinfo {pages} {195107} (\bibinfo {year} {2006})}\BibitemShut
  {NoStop}%
\bibitem [{\citenamefont {Jain}\ \emph
  {et~al.}(2011{\natexlab{b}})\citenamefont {Jain}, \citenamefont {Hautier},
  \citenamefont {Ong}, \citenamefont {Moore}, \citenamefont {Fischer},
  \citenamefont {Persson},\ and\ \citenamefont {Ceder}}]{Jain2011a}%
  \BibitemOpen
  \bibfield  {author} {\bibinfo {author} {\bibfnamefont {A.}~\bibnamefont
  {Jain}}, \bibinfo {author} {\bibfnamefont {G.}~\bibnamefont {Hautier}},
  \bibinfo {author} {\bibfnamefont {S.~P.}\ \bibnamefont {Ong}}, \bibinfo
  {author} {\bibfnamefont {C.~J.}\ \bibnamefont {Moore}}, \bibinfo {author}
  {\bibfnamefont {C.~C.}\ \bibnamefont {Fischer}}, \bibinfo {author}
  {\bibfnamefont {K.~A.}\ \bibnamefont {Persson}}, \ and\ \bibinfo {author}
  {\bibfnamefont {G.}~\bibnamefont {Ceder}},\ }\href@noop {} {\bibfield
  {journal} {\bibinfo  {journal} {Physical Review B - Condensed Matter and
  Materials Physics}\ }\textbf {\bibinfo {volume} {84}},\ \bibinfo {pages}
  {045115} (\bibinfo {year} {2011}{\natexlab{b}})}\BibitemShut {NoStop}%
\bibitem [{\citenamefont {Hinuma}\ \emph {et~al.}(2017)\citenamefont {Hinuma},
  \citenamefont {Pizzi}, \citenamefont {Kumagai}, \citenamefont {Oba},\ and\
  \citenamefont {Tanaka}}]{Hinuma2017}%
  \BibitemOpen
  \bibfield  {author} {\bibinfo {author} {\bibfnamefont {Y.}~\bibnamefont
  {Hinuma}}, \bibinfo {author} {\bibfnamefont {G.}~\bibnamefont {Pizzi}},
  \bibinfo {author} {\bibfnamefont {Y.}~\bibnamefont {Kumagai}}, \bibinfo
  {author} {\bibfnamefont {F.}~\bibnamefont {Oba}}, \ and\ \bibinfo {author}
  {\bibfnamefont {I.}~\bibnamefont {Tanaka}},\ }\href {\doibase
  10.1016/j.commatsci.2016.10.015} {\bibfield  {journal} {\bibinfo  {journal}
  {Computational Materials Science}\ }\textbf {\bibinfo {volume} {128}},\
  \bibinfo {pages} {140} (\bibinfo {year} {2017})}\BibitemShut {NoStop}%
\bibitem [{\citenamefont {Herath}\ \emph {et~al.}(2020)\citenamefont {Herath},
  \citenamefont {Tavadze}, \citenamefont {He}, \citenamefont {Bousquet},
  \citenamefont {Singh}, \citenamefont {Mu{\~{n}}oz},\ and\ \citenamefont
  {Romero}}]{Herath2020}%
  \BibitemOpen
  \bibfield  {author} {\bibinfo {author} {\bibfnamefont {U.}~\bibnamefont
  {Herath}}, \bibinfo {author} {\bibfnamefont {P.}~\bibnamefont {Tavadze}},
  \bibinfo {author} {\bibfnamefont {X.}~\bibnamefont {He}}, \bibinfo {author}
  {\bibfnamefont {E.}~\bibnamefont {Bousquet}}, \bibinfo {author}
  {\bibfnamefont {S.}~\bibnamefont {Singh}}, \bibinfo {author} {\bibfnamefont
  {F.}~\bibnamefont {Mu{\~{n}}oz}}, \ and\ \bibinfo {author} {\bibfnamefont
  {A.~H.}\ \bibnamefont {Romero}},\ }\href@noop {} {\bibfield  {journal}
  {\bibinfo  {journal} {Computer Physics Communications}\ }\textbf {\bibinfo
  {volume} {251}},\ \bibinfo {pages} {107080} (\bibinfo {year}
  {2020})}\BibitemShut {NoStop}%
\bibitem [{\citenamefont {Heyd}\ \emph {et~al.}(2005)\citenamefont {Heyd},
  \citenamefont {Scuseria}, \citenamefont {Ernzerhof}, \citenamefont {Heyd},
  \citenamefont {Scuseria},\ and\ \citenamefont {Ernzerhof}}]{Heyd2003}%
  \BibitemOpen
  \bibfield  {author} {\bibinfo {author} {\bibfnamefont {J.}~\bibnamefont
  {Heyd}}, \bibinfo {author} {\bibfnamefont {G.~E.}\ \bibnamefont {Scuseria}},
  \bibinfo {author} {\bibfnamefont {M.}~\bibnamefont {Ernzerhof}}, \bibinfo
  {author} {\bibfnamefont {J.}~\bibnamefont {Heyd}}, \bibinfo {author}
  {\bibfnamefont {G.~E.}\ \bibnamefont {Scuseria}}, \ and\ \bibinfo {author}
  {\bibfnamefont {M.}~\bibnamefont {Ernzerhof}},\ }\href@noop {} {\bibfield
  {journal} {\bibinfo  {journal} {The Journal of Chemical Physics}\ }\textbf
  {\bibinfo {volume} {118}},\ \bibinfo {pages} {8207} (\bibinfo {year}
  {2005})}\BibitemShut {NoStop}%
\bibitem [{\citenamefont {Heyd}\ \emph {et~al.}(2006)\citenamefont {Heyd},
  \citenamefont {Scuseria},\ and\ \citenamefont {Ernzerhof}}]{Heyd2006}%
  \BibitemOpen
  \bibfield  {author} {\bibinfo {author} {\bibfnamefont {J.}~\bibnamefont
  {Heyd}}, \bibinfo {author} {\bibfnamefont {G.~E.}\ \bibnamefont {Scuseria}},
  \ and\ \bibinfo {author} {\bibfnamefont {M.}~\bibnamefont {Ernzerhof}},\
  }\href@noop {} {\bibfield  {journal} {\bibinfo  {journal} {Journal of
  Chemical Physics}\ }\textbf {\bibinfo {volume} {124}},\ \bibinfo {pages}
  {219906} (\bibinfo {year} {2006})}\BibitemShut {NoStop}%
\bibitem [{\citenamefont {Togo}\ and\ \citenamefont {Tanaka}(2015)}]{Togo2015}%
  \BibitemOpen
  \bibfield  {author} {\bibinfo {author} {\bibfnamefont {A.}~\bibnamefont
  {Togo}}\ and\ \bibinfo {author} {\bibfnamefont {I.}~\bibnamefont {Tanaka}},\
  }\href@noop {} {\bibfield  {journal} {\bibinfo  {journal} {Scripta
  Materialia}\ }\textbf {\bibinfo {volume} {108}},\ \bibinfo {pages} {1}
  (\bibinfo {year} {2015})},\ \Eprint {http://arxiv.org/abs/1506.08498}
  {1506.08498} \BibitemShut {NoStop}%
\bibitem [{\citenamefont {King-Smith}\ and\ \citenamefont
  {Vanderbilt}(1993)}]{King-Smith1993}%
  \BibitemOpen
  \bibfield  {author} {\bibinfo {author} {\bibfnamefont {R.~D.}\ \bibnamefont
  {King-Smith}}\ and\ \bibinfo {author} {\bibfnamefont {D.}~\bibnamefont
  {Vanderbilt}},\ }\href {\doibase 10.1103/PhysRevB.47.1651} {\bibfield
  {journal} {\bibinfo  {journal} {Physical Review B}\ }\textbf {\bibinfo
  {volume} {47}},\ \bibinfo {pages} {1651} (\bibinfo {year}
  {1993})}\BibitemShut {NoStop}%
\bibitem [{\citenamefont {Zagorac}\ \emph {et~al.}(2019)\citenamefont
  {Zagorac}, \citenamefont {Muller}, \citenamefont {Ruehl}, \citenamefont
  {Zagorac},\ and\ \citenamefont {Rehme}}]{Zagorac2019}%
  \BibitemOpen
  \bibfield  {author} {\bibinfo {author} {\bibfnamefont {D.}~\bibnamefont
  {Zagorac}}, \bibinfo {author} {\bibfnamefont {H.}~\bibnamefont {Muller}},
  \bibinfo {author} {\bibfnamefont {S.}~\bibnamefont {Ruehl}}, \bibinfo
  {author} {\bibfnamefont {J.}~\bibnamefont {Zagorac}}, \ and\ \bibinfo
  {author} {\bibfnamefont {S.}~\bibnamefont {Rehme}},\ }\href {\doibase
  10.1107/S160057671900997X} {\bibfield  {journal} {\bibinfo  {journal}
  {Journal of Applied Crystallography}\ }\textbf {\bibinfo {volume} {52}},\
  \bibinfo {pages} {918} (\bibinfo {year} {2019})}\BibitemShut {NoStop}%
\bibitem [{\citenamefont {Scheidgen}\ \emph {et~al.}(2023)\citenamefont
  {Scheidgen}, \citenamefont {Himanen}, \citenamefont {Ladines}, \citenamefont
  {Sikter}, \citenamefont {Nakhaee}, \citenamefont {Fekete}, \citenamefont
  {Chang}, \citenamefont {Golparvar}, \citenamefont {M{\'{a}}rquez},
  \citenamefont {Brockhauser}, \citenamefont {Br{\"{u}}ckner}, \citenamefont
  {Ghiringhelli}, \citenamefont {Dietrich}, \citenamefont {Lehmberg},
  \citenamefont {Denell}, \citenamefont {Albino}, \citenamefont
  {N{\"{a}}sstr{\"{o}}m}, \citenamefont {Shabih}, \citenamefont {Dobener},
  \citenamefont {K{\"{u}}hbach}, \citenamefont {Mozumder}, \citenamefont
  {Rudzinski}, \citenamefont {Daelman}, \citenamefont {Pizarro}, \citenamefont
  {Kuban}, \citenamefont {Salazar}, \citenamefont {Ondra{\v{c}}ka},
  \citenamefont {Bungartz},\ and\ \citenamefont {Draxl}}]{Scheidgen2023}%
  \BibitemOpen
  \bibfield  {author} {\bibinfo {author} {\bibfnamefont {M.}~\bibnamefont
  {Scheidgen}}, \bibinfo {author} {\bibfnamefont {L.}~\bibnamefont {Himanen}},
  \bibinfo {author} {\bibfnamefont {A.~N.}\ \bibnamefont {Ladines}}, \bibinfo
  {author} {\bibfnamefont {D.}~\bibnamefont {Sikter}}, \bibinfo {author}
  {\bibfnamefont {M.}~\bibnamefont {Nakhaee}}, \bibinfo {author} {\bibfnamefont
  {{\'{A}}.}~\bibnamefont {Fekete}}, \bibinfo {author} {\bibfnamefont
  {T.}~\bibnamefont {Chang}}, \bibinfo {author} {\bibfnamefont
  {A.}~\bibnamefont {Golparvar}}, \bibinfo {author} {\bibfnamefont {J.~A.}\
  \bibnamefont {M{\'{a}}rquez}}, \bibinfo {author} {\bibfnamefont
  {S.}~\bibnamefont {Brockhauser}}, \bibinfo {author} {\bibfnamefont
  {S.}~\bibnamefont {Br{\"{u}}ckner}}, \bibinfo {author} {\bibfnamefont
  {L.~M.}\ \bibnamefont {Ghiringhelli}}, \bibinfo {author} {\bibfnamefont
  {F.}~\bibnamefont {Dietrich}}, \bibinfo {author} {\bibfnamefont
  {D.}~\bibnamefont {Lehmberg}}, \bibinfo {author} {\bibfnamefont
  {T.}~\bibnamefont {Denell}}, \bibinfo {author} {\bibfnamefont
  {A.}~\bibnamefont {Albino}}, \bibinfo {author} {\bibfnamefont
  {H.}~\bibnamefont {N{\"{a}}sstr{\"{o}}m}}, \bibinfo {author} {\bibfnamefont
  {S.}~\bibnamefont {Shabih}}, \bibinfo {author} {\bibfnamefont
  {F.}~\bibnamefont {Dobener}}, \bibinfo {author} {\bibfnamefont
  {M.}~\bibnamefont {K{\"{u}}hbach}}, \bibinfo {author} {\bibfnamefont
  {R.}~\bibnamefont {Mozumder}}, \bibinfo {author} {\bibfnamefont {J.~F.}\
  \bibnamefont {Rudzinski}}, \bibinfo {author} {\bibfnamefont {N.}~\bibnamefont
  {Daelman}}, \bibinfo {author} {\bibfnamefont {J.~M.}\ \bibnamefont
  {Pizarro}}, \bibinfo {author} {\bibfnamefont {M.}~\bibnamefont {Kuban}},
  \bibinfo {author} {\bibfnamefont {C.}~\bibnamefont {Salazar}}, \bibinfo
  {author} {\bibfnamefont {P.}~\bibnamefont {Ondra{\v{c}}ka}}, \bibinfo
  {author} {\bibfnamefont {H.-J.}\ \bibnamefont {Bungartz}}, \ and\ \bibinfo
  {author} {\bibfnamefont {C.}~\bibnamefont {Draxl}},\ }\href {\doibase
  10.21105/joss.05388} {\bibfield  {journal} {\bibinfo  {journal} {Journal of
  Open Source Software}\ }\textbf {\bibinfo {volume} {8}},\ \bibinfo {pages}
  {5388} (\bibinfo {year} {2023})}\BibitemShut {NoStop}%
\bibitem [{\citenamefont {Wines}\ \emph {et~al.}(2023)\citenamefont {Wines},
  \citenamefont {Gurunathan}, \citenamefont {Garrity}, \citenamefont {DeCost},
  \citenamefont {Biacchi}, \citenamefont {Tavazza},\ and\ \citenamefont
  {Choudhary}}]{Wines2023}%
  \BibitemOpen
  \bibfield  {author} {\bibinfo {author} {\bibfnamefont {D.}~\bibnamefont
  {Wines}}, \bibinfo {author} {\bibfnamefont {R.}~\bibnamefont {Gurunathan}},
  \bibinfo {author} {\bibfnamefont {K.~F.}\ \bibnamefont {Garrity}}, \bibinfo
  {author} {\bibfnamefont {B.}~\bibnamefont {DeCost}}, \bibinfo {author}
  {\bibfnamefont {A.~J.}\ \bibnamefont {Biacchi}}, \bibinfo {author}
  {\bibfnamefont {F.}~\bibnamefont {Tavazza}}, \ and\ \bibinfo {author}
  {\bibfnamefont {K.}~\bibnamefont {Choudhary}},\ }\href@noop {} {\bibfield
  {journal} {\bibinfo  {journal} {Applied Physics Reviews}\ }\textbf {\bibinfo
  {volume} {10}} (\bibinfo {year} {2023})}\BibitemShut {NoStop}%
\bibitem [{\citenamefont {Talirz}\ \emph {et~al.}(2020)\citenamefont {Talirz},
  \citenamefont {Kumbhar}, \citenamefont {Passaro}, \citenamefont {Yakutovich},
  \citenamefont {Granata}, \citenamefont {Gargiulo}, \citenamefont {Borelli},
  \citenamefont {Uhrin}, \citenamefont {Huber}, \citenamefont {Zoupanos},
  \citenamefont {Adorf}, \citenamefont {Andersen}, \citenamefont
  {Sch{\"{u}}tt}, \citenamefont {Pignedoli}, \citenamefont {Passerone},
  \citenamefont {VandeVondele}, \citenamefont {Schulthess}, \citenamefont
  {Smit}, \citenamefont {Pizzi},\ and\ \citenamefont {Marzari}}]{Talirz2020}%
  \BibitemOpen
  \bibfield  {author} {\bibinfo {author} {\bibfnamefont {L.}~\bibnamefont
  {Talirz}}, \bibinfo {author} {\bibfnamefont {S.}~\bibnamefont {Kumbhar}},
  \bibinfo {author} {\bibfnamefont {E.}~\bibnamefont {Passaro}}, \bibinfo
  {author} {\bibfnamefont {A.~V.}\ \bibnamefont {Yakutovich}}, \bibinfo
  {author} {\bibfnamefont {V.}~\bibnamefont {Granata}}, \bibinfo {author}
  {\bibfnamefont {F.}~\bibnamefont {Gargiulo}}, \bibinfo {author}
  {\bibfnamefont {M.}~\bibnamefont {Borelli}}, \bibinfo {author} {\bibfnamefont
  {M.}~\bibnamefont {Uhrin}}, \bibinfo {author} {\bibfnamefont {S.~P.}\
  \bibnamefont {Huber}}, \bibinfo {author} {\bibfnamefont {S.}~\bibnamefont
  {Zoupanos}}, \bibinfo {author} {\bibfnamefont {C.~S.}\ \bibnamefont {Adorf}},
  \bibinfo {author} {\bibfnamefont {C.~W.}\ \bibnamefont {Andersen}}, \bibinfo
  {author} {\bibfnamefont {O.}~\bibnamefont {Sch{\"{u}}tt}}, \bibinfo {author}
  {\bibfnamefont {C.~A.}\ \bibnamefont {Pignedoli}}, \bibinfo {author}
  {\bibfnamefont {D.}~\bibnamefont {Passerone}}, \bibinfo {author}
  {\bibfnamefont {J.}~\bibnamefont {VandeVondele}}, \bibinfo {author}
  {\bibfnamefont {T.~C.}\ \bibnamefont {Schulthess}}, \bibinfo {author}
  {\bibfnamefont {B.}~\bibnamefont {Smit}}, \bibinfo {author} {\bibfnamefont
  {G.}~\bibnamefont {Pizzi}}, \ and\ \bibinfo {author} {\bibfnamefont
  {N.}~\bibnamefont {Marzari}},\ }\href@noop {} {\bibfield  {journal} {\bibinfo
   {journal} {Scientific Data}\ }\textbf {\bibinfo {volume} {7}},\ \bibinfo
  {pages} {1} (\bibinfo {year} {2020})}\BibitemShut {NoStop}%
\bibitem [{\citenamefont {Didisheim}\ \emph {et~al.}(1986)\citenamefont
  {Didisheim}, \citenamefont {McMullan},\ and\ \citenamefont
  {Wuensch}}]{Didisheim1986}%
  \BibitemOpen
  \bibfield  {author} {\bibinfo {author} {\bibfnamefont {J.~J.}\ \bibnamefont
  {Didisheim}}, \bibinfo {author} {\bibfnamefont {R.~K.}\ \bibnamefont
  {McMullan}}, \ and\ \bibinfo {author} {\bibfnamefont {B.~J.}\ \bibnamefont
  {Wuensch}},\ }\href@noop {} {\bibfield  {journal} {\bibinfo  {journal} {Solid
  State Ionics}\ }\textbf {\bibinfo {volume} {18-19}},\ \bibinfo {pages} {1150}
  (\bibinfo {year} {1986})}\BibitemShut {NoStop}%
\bibitem [{\citenamefont {Christensen}\ \emph {et~al.}(1964)\citenamefont
  {Christensen}, \citenamefont {Gronbek},\ and\ \citenamefont
  {Rasmussen}}]{Christensen1964}%
  \BibitemOpen
  \bibfield  {author} {\bibinfo {author} {\bibfnamefont {A.~N.}\ \bibnamefont
  {Christensen}}, \bibinfo {author} {\bibfnamefont {R.}~\bibnamefont
  {Gronbek}}, \ and\ \bibinfo {author} {\bibfnamefont {S.~E.}\ \bibnamefont
  {Rasmussen}},\ }\href@noop {} {\bibfield  {journal} {\bibinfo  {journal}
  {Acta Chemica Scandinavica}\ }\textbf {\bibinfo {volume} {18}},\ \bibinfo
  {pages} {1261} (\bibinfo {year} {1964})}\BibitemShut {NoStop}%
\bibitem [{\citenamefont {Lehmann}\ \emph {et~al.}(1970)\citenamefont
  {Lehmann}, \citenamefont {Larsen}, \citenamefont {Poulsen}, \citenamefont
  {Christensen}, \citenamefont {Rasmussen}, \citenamefont {Sunde},\ and\
  \citenamefont {S{\o}rensen}}]{Lehmann1970}%
  \BibitemOpen
  \bibfield  {author} {\bibinfo {author} {\bibfnamefont {M.~S.}\ \bibnamefont
  {Lehmann}}, \bibinfo {author} {\bibfnamefont {F.~K.}\ \bibnamefont {Larsen}},
  \bibinfo {author} {\bibfnamefont {F.~R.}\ \bibnamefont {Poulsen}}, \bibinfo
  {author} {\bibfnamefont {A.~N.}\ \bibnamefont {Christensen}}, \bibinfo
  {author} {\bibfnamefont {S.~E.}\ \bibnamefont {Rasmussen}}, \bibinfo {author}
  {\bibfnamefont {E.}~\bibnamefont {Sunde}}, \ and\ \bibinfo {author}
  {\bibfnamefont {N.~A.}\ \bibnamefont {S{\o}rensen}},\ }\href {\doibase
  10.3891/acta.chem.scand.24-1662} {\bibfield  {journal} {\bibinfo  {journal}
  {Acta Chemica Scandinavica}\ }\textbf {\bibinfo {volume} {24}},\ \bibinfo
  {pages} {1662} (\bibinfo {year} {1970})}\BibitemShut {NoStop}%
\bibitem [{\citenamefont {Endo}\ \emph {et~al.}(1989)\citenamefont {Endo},
  \citenamefont {Chino}, \citenamefont {Tsuboi},\ and\ \citenamefont
  {Koto}}]{Endo1989}%
  \BibitemOpen
  \bibfield  {author} {\bibinfo {author} {\bibfnamefont {S.}~\bibnamefont
  {Endo}}, \bibinfo {author} {\bibfnamefont {T.}~\bibnamefont {Chino}},
  \bibinfo {author} {\bibfnamefont {S.}~\bibnamefont {Tsuboi}}, \ and\ \bibinfo
  {author} {\bibfnamefont {K.}~\bibnamefont {Koto}},\ }\href {\doibase
  10.1038/340452a0} {\bibfield  {journal} {\bibinfo  {journal} {Nature}\
  }\textbf {\bibinfo {volume} {340}},\ \bibinfo {pages} {452} (\bibinfo {year}
  {1989})}\BibitemShut {NoStop}%
\bibitem [{\citenamefont {Miyoshi}\ \emph {et~al.}(2011)\citenamefont
  {Miyoshi}, \citenamefont {Mashiyama}, \citenamefont {Asahi}, \citenamefont
  {Kimura},\ and\ \citenamefont {Noda}}]{Miyoshi2011}%
  \BibitemOpen
  \bibfield  {author} {\bibinfo {author} {\bibfnamefont {T.}~\bibnamefont
  {Miyoshi}}, \bibinfo {author} {\bibfnamefont {H.}~\bibnamefont {Mashiyama}},
  \bibinfo {author} {\bibfnamefont {T.}~\bibnamefont {Asahi}}, \bibinfo
  {author} {\bibfnamefont {H.}~\bibnamefont {Kimura}}, \ and\ \bibinfo {author}
  {\bibfnamefont {Y.}~\bibnamefont {Noda}},\ }\href {\doibase
  10.1143/JPSJ.80.044709} {\bibfield  {journal} {\bibinfo  {journal} {Journal
  of the Physical Society of Japan}\ }\textbf {\bibinfo {volume} {80}},\
  \bibinfo {pages} {044709} (\bibinfo {year} {2011})}\BibitemShut {NoStop}%
\bibitem [{\citenamefont {Sueno}\ \emph {et~al.}(1973)\citenamefont {Sueno},
  \citenamefont {Clark}, \citenamefont {Papike},\ and\ \citenamefont
  {Konnert}}]{Sueno1973}%
  \BibitemOpen
  \bibfield  {author} {\bibinfo {author} {\bibfnamefont {S.}~\bibnamefont
  {Sueno}}, \bibinfo {author} {\bibfnamefont {J.~J.}\ \bibnamefont {Clark}},
  \bibinfo {author} {\bibfnamefont {J.~J.}\ \bibnamefont {Papike}}, \ and\
  \bibinfo {author} {\bibfnamefont {J.~A.}\ \bibnamefont {Konnert}},\
  }\href@noop {} {\bibfield  {journal} {\bibinfo  {journal} {American
  Mineralogist}\ }\textbf {\bibinfo {volume} {58}},\ \bibinfo {pages} {691}
  (\bibinfo {year} {1973})}\BibitemShut {NoStop}%
\bibitem [{\citenamefont {Wang}\ \emph {et~al.}(2018)\citenamefont {Wang},
  \citenamefont {Gaskell}, \citenamefont {Dopita}, \citenamefont {Kriegner},
  \citenamefont {Tasneem}, \citenamefont {Mack}, \citenamefont {Mukherjee},
  \citenamefont {Karim},\ and\ \citenamefont {Khan}}]{Wang2018}%
  \BibitemOpen
  \bibfield  {author} {\bibinfo {author} {\bibfnamefont {Z.}~\bibnamefont
  {Wang}}, \bibinfo {author} {\bibfnamefont {A.~A.}\ \bibnamefont {Gaskell}},
  \bibinfo {author} {\bibfnamefont {M.}~\bibnamefont {Dopita}}, \bibinfo
  {author} {\bibfnamefont {D.}~\bibnamefont {Kriegner}}, \bibinfo {author}
  {\bibfnamefont {N.}~\bibnamefont {Tasneem}}, \bibinfo {author} {\bibfnamefont
  {J.}~\bibnamefont {Mack}}, \bibinfo {author} {\bibfnamefont {N.}~\bibnamefont
  {Mukherjee}}, \bibinfo {author} {\bibfnamefont {Z.}~\bibnamefont {Karim}}, \
  and\ \bibinfo {author} {\bibfnamefont {A.~I.}\ \bibnamefont {Khan}},\
  }\href@noop {} {\bibfield  {journal} {\bibinfo  {journal} {Applied Physics
  Letters}\ }\textbf {\bibinfo {volume} {112}},\ \bibinfo {pages} {222902}
  (\bibinfo {year} {2018})}\BibitemShut {NoStop}%
\bibitem [{\citenamefont {Mattauch}\ \emph {et~al.}(2004)\citenamefont
  {Mattauch}, \citenamefont {Heger},\ and\ \citenamefont
  {Michel}}]{Mattauch2004}%
  \BibitemOpen
  \bibfield  {author} {\bibinfo {author} {\bibfnamefont {S.}~\bibnamefont
  {Mattauch}}, \bibinfo {author} {\bibfnamefont {G.}~\bibnamefont {Heger}}, \
  and\ \bibinfo {author} {\bibfnamefont {K.~H.}\ \bibnamefont {Michel}},\
  }\href {\doibase 10.1002/crat.200410289} {\bibfield  {journal} {\bibinfo
  {journal} {Crystal Research and Technology}\ }\textbf {\bibinfo {volume}
  {39}},\ \bibinfo {pages} {1027} (\bibinfo {year} {2004})}\BibitemShut
  {NoStop}%
\bibitem [{\citenamefont {Momma}\ and\ \citenamefont
  {Izumi}(2008)}]{Momma2008}%
  \BibitemOpen
  \bibfield  {author} {\bibinfo {author} {\bibfnamefont {K.}~\bibnamefont
  {Momma}}\ and\ \bibinfo {author} {\bibfnamefont {F.}~\bibnamefont {Izumi}},\
  }\href {\doibase 10.1107/S0021889808012016} {\bibfield  {journal} {\bibinfo
  {journal} {Applied Crystallography}\ }\textbf {\bibinfo {volume} {41}},\
  \bibinfo {pages} {653} (\bibinfo {year} {2008})}\BibitemShut {NoStop}%
\bibitem [{\citenamefont {Lines}\ and\ \citenamefont
  {Glass}(1977)}]{linesbook}%
  \BibitemOpen
  \bibfield  {author} {\bibinfo {author} {\bibfnamefont {M.~E.}\ \bibnamefont
  {Lines}}\ and\ \bibinfo {author} {\bibfnamefont {A.~M.}\ \bibnamefont
  {Glass}},\ }\href@noop {} { {\bibinfo {title} {{Principles and
  Applications of Ferroelectrics and Related Materials}}}}\ (\bibinfo
  {publisher} {Clarendon Press Oxford University},\ \bibinfo {year}
  {1977})\BibitemShut {NoStop}%
\bibitem [{\citenamefont {Cancarevic}\ \emph {et~al.}(2007)\citenamefont
  {Cancarevic}, \citenamefont {Schon},\ and\ \citenamefont
  {Jansen}}]{Cancarevic2007}%
  \BibitemOpen
  \bibfield  {author} {\bibinfo {author} {\bibfnamefont {Z.}~\bibnamefont
  {Cancarevic}}, \bibinfo {author} {\bibfnamefont {J.~C.}\ \bibnamefont
  {Schon}}, \ and\ \bibinfo {author} {\bibfnamefont {M.}~\bibnamefont
  {Jansen}},\ }\href {\doibase 10.1002/chem.200601637} {\bibfield  {journal}
  {\bibinfo  {journal} {Chem. Eur. J.}\ }\textbf {\bibinfo {volume} {13}},\
  \bibinfo {pages} {7330} (\bibinfo {year} {2007})}\BibitemShut {NoStop}%
\bibitem [{\citenamefont {Mori-S{\'{a}}nchez}\ \emph
  {et~al.}(2008)\citenamefont {Mori-S{\'{a}}nchez}, \citenamefont {Cohen},\
  and\ \citenamefont {Yang}}]{Mori-Sanchez2008}%
  \BibitemOpen
  \bibfield  {author} {\bibinfo {author} {\bibfnamefont {P.}~\bibnamefont
  {Mori-S{\'{a}}nchez}}, \bibinfo {author} {\bibfnamefont {A.~J.}\ \bibnamefont
  {Cohen}}, \ and\ \bibinfo {author} {\bibfnamefont {W.}~\bibnamefont {Yang}},\
  }\href {\doibase 10.1103/PhysRevLett.100.146401} {\bibfield  {journal}
  {\bibinfo  {journal} {Physical Review Letters}\ }\textbf {\bibinfo {volume}
  {100}},\ \bibinfo {pages} {146401} (\bibinfo {year} {2008})}\BibitemShut
  {NoStop}%
\bibitem [{\citenamefont {Krach}\ \emph {et~al.}(2023)\citenamefont {Krach},
  \citenamefont {Forero-Correa}, \citenamefont {Biega}, \citenamefont
  {Reyes-Lillo},\ and\ \citenamefont {Leppert}}]{Krach2023}%
  \BibitemOpen
  \bibfield  {author} {\bibinfo {author} {\bibfnamefont {S.}~\bibnamefont
  {Krach}}, \bibinfo {author} {\bibfnamefont {N.}~\bibnamefont
  {Forero-Correa}}, \bibinfo {author} {\bibfnamefont {R.~I.}\ \bibnamefont
  {Biega}}, \bibinfo {author} {\bibfnamefont {S.~E.}\ \bibnamefont
  {Reyes-Lillo}}, \ and\ \bibinfo {author} {\bibfnamefont {L.}~\bibnamefont
  {Leppert}},\ }\href@noop {} {\bibfield  {journal} {\bibinfo  {journal}
  {Journal of Physics Condensed Matter}\ }\textbf {\bibinfo {volume} {35}},\
  \bibinfo {pages} {174001} (\bibinfo {year} {2023})}\BibitemShut {NoStop}%
\bibitem [{\citenamefont {Komatsu}\ \emph {et~al.}(2006)\citenamefont
  {Komatsu}, \citenamefont {Kuribayashi}, \citenamefont {Sano}, \citenamefont
  {Ohtani},\ and\ \citenamefont {Kudoh}}]{Komatsu2006}%
  \BibitemOpen
  \bibfield  {author} {\bibinfo {author} {\bibfnamefont {K.}~\bibnamefont
  {Komatsu}}, \bibinfo {author} {\bibfnamefont {T.}~\bibnamefont
  {Kuribayashi}}, \bibinfo {author} {\bibfnamefont {A.}~\bibnamefont {Sano}},
  \bibinfo {author} {\bibfnamefont {E.}~\bibnamefont {Ohtani}}, \ and\ \bibinfo
  {author} {\bibfnamefont {Y.}~\bibnamefont {Kudoh}},\ }\href {\doibase
  10.1107/S160053680603916X} {\bibfield  {journal} {\bibinfo  {journal} {Acta
  Crystallographica Section E: Structure Reports Online}\ }\textbf {\bibinfo
  {volume} {62}},\ \bibinfo {pages} {216} (\bibinfo {year} {2006})}\BibitemShut
  {NoStop}%
\bibitem [{\citenamefont {{Van Aken}}\ \emph {et~al.}(2004)\citenamefont {{Van
  Aken}}, \citenamefont {Palstra}, \citenamefont {Filippetti},\ and\
  \citenamefont {Spaldin}}]{Aken2004}%
  \BibitemOpen
  \bibfield  {author} {\bibinfo {author} {\bibfnamefont {B.~B.}\ \bibnamefont
  {{Van Aken}}}, \bibinfo {author} {\bibfnamefont {T.~T.~M.}\ \bibnamefont
  {Palstra}}, \bibinfo {author} {\bibfnamefont {A.}~\bibnamefont {Filippetti}},
  \ and\ \bibinfo {author} {\bibfnamefont {N.~A.}\ \bibnamefont {Spaldin}},\
  }\href@noop {} {\bibfield  {journal} {\bibinfo  {journal} {Nature materials}\
  }\textbf {\bibinfo {volume} {3}},\ \bibinfo {pages} {164} (\bibinfo {year}
  {2004})}\BibitemShut {NoStop}%
\bibitem [{\citenamefont {Campa-Molina}\ \emph {et~al.}(2006)\citenamefont
  {Campa-Molina}, \citenamefont {Ulloa-God{\'{i}}nez}, \citenamefont {Barrera},
  \citenamefont {Bucio},\ and\ \citenamefont {Mata}}]{Campa-Molina2006}%
  \BibitemOpen
  \bibfield  {author} {\bibinfo {author} {\bibfnamefont {J.}~\bibnamefont
  {Campa-Molina}}, \bibinfo {author} {\bibfnamefont {S.}~\bibnamefont
  {Ulloa-God{\'{i}}nez}}, \bibinfo {author} {\bibfnamefont {A.}~\bibnamefont
  {Barrera}}, \bibinfo {author} {\bibfnamefont {L.}~\bibnamefont {Bucio}}, \
  and\ \bibinfo {author} {\bibfnamefont {J.}~\bibnamefont {Mata}},\ }\href
  {\doibase 10.1088/0953-8984/18/20/007} {\bibfield  {journal} {\bibinfo
  {journal} {Journal of Physics Condensed Matter}\ }\textbf {\bibinfo {volume}
  {18}},\ \bibinfo {pages} {4827} (\bibinfo {year} {2006})}\BibitemShut
  {NoStop}%
\end{thebibliography}

%

\end{document}